\documentclass[twocolumn,showpacs,preprintnumbers,superscriptaddress,amsmath,amssymb,aps,pre]{revtex4}

\usepackage{graphicx}
\usepackage{dcolumn}
\usepackage{bm}
\usepackage{color}

\begin{document}


\title{Recurrence interval analysis of trading volumes}

\author{Fei Ren}
 \email{fren@ecust.edu.cn}
 \affiliation{School of Business, East China University of Science and Technology, Shanghai 200237, China} %
 \affiliation{Research Center for Econophysics, East China University of Science and Technology, Shanghai 200237, China} %

\author{Wei-Xing Zhou}
 \email{wxzhou@ecust.edu.cn}
 \affiliation{School of Business, East China University of Science and Technology, Shanghai 200237, China} %
 \affiliation{Research Center for Econophysics, East China University of Science and Technology, Shanghai 200237, China} %
 \affiliation{School of Science, East China University of Science and Technology, Shanghai 200237, China} %

\date{\today}

\begin{abstract}
We study the statistical properties of the recurrence intervals $\tau$ between successive trading volumes exceeding a certain
threshold $q$. The recurrence interval analysis is carried out for the $20$ liquid Chinese stocks covering a period from January 2000 to May 2009, and two Chinese indices from January 2003 to April 2009. Similar to the recurrence interval distribution of the price returns, the tail of the recurrence interval distribution of the trading volumes follows a power-law scaling, and the results are verified by the goodness-of-fit tests using the Kolmogorov-Smirnov (KS) statistic, the weighted KS statistic and the Cram{\'{e}}r-von Mises criterion. The measurements of the conditional probability distribution and the detrended fluctuation function show that both short-term and long-term memory effects exist in the recurrence intervals between trading volumes. We further study the relationship between trading volumes and price returns based on the recurrence interval analysis method. It is found that large trading volumes are more likely to occur following large price returns, and the comovement between trading volumes and price returns is more pronounced for large trading volumes.
\end{abstract}

\pacs{05.45.Tp, 89.75.Da, 89.65.Gh}

\maketitle


\section{Introduction}

Numerous studies have been conducted to
better understand the dynamic behavior of financial
markets. Most of the studies mainly focus on the statistical properties of the independent
events occurring at discrete time steps. In recent years, the waiting time between consecutive
events has drawn much attention, for instance, the intertrade duration between
two consecutive trades \cite{Scalas-Gorenflo-Luckock-Mainardi-Mantelli-Raberto-2004-QF,Ivanov-Yuen-Podobnik-Lee-2004-PRE,Sazuka-2007-PA,Jiang-Chen-Zhou-2008-PA},
the duration time that the price or volatility
keeps below or above its initial value \cite{Zheng-2002-MPLB,Ren-Zheng-2003-PLA,Ren-Zheng-Lin-Wen-Trimper-2005-PA},
and the waiting time that the price return first exceeds a predefined level
\cite{Simonsen-Jensen-Johansen-2002-EPJB,Jensen-Johansen-Simonsen-2003-IJMPC,Jensen-Johansen-Simonsen-2003-PA,ZaluskaKotur-Karpio-Orlowski-2006-APPB,Karpio-ZaluskaKotur-Orlowski-2007-PA,Zhou-Yuan-2005-PA}.
To the best of our knowledge, the concepts raised above mainly concern the waiting
time between normal events in financial markets, i.e., the measurement of the events themselves
has not been considered.

The study of recurrence intervals between extreme events in natural records, e.g. floods, temperatures and earthquakes
\cite{Bunde-Eichner-Havlin-Kantelhardt-2003-PA,Bunde-Eichner-Havlin-Kantelhardt-2004-PA,Bunde-Eichner-Kantelhardt-Havlin-2005-PRL,Saichev-Sornette-2006-PRL}, has become a focus of research in climatic cataclysm, and has been widely studied by many scientists. It offers an inspiration that the analysis of recurrence intervals between extreme events with large price fluctuations may offer an
alternative way of learning the dynamic behavior of financial markets. Rather than studying the
waiting time between ordinary events, the recurrence interval analysis is more concerned
about the duration time between events exceeding a certain threshold level. The recurrence
intervals between volatilities exceeding a threshold $q$ have been carefully studied, and it gets
to two primarily divergent opinions about the recurrence interval distribution: power-law and stretched exponential
distributions are observed respectively for a variety of representative financial records \cite{Kaizoji-Kaizoji-2004a-PA,Yamasaki-Muchnik-Havlin-Bunde-Stanley-2005-PNAS,Wang-Yamasaki-Havlin-Stanley-2006-PRE,Lee-Lee-Rikvold-2006-JKPS,Wang-Weber-Yamasaki-Havlin-Stanley-2007-EPJB,VodenskaChitkushev-Wang-Weber-Yamasaki-Havlin-Stanley-2008-EPJB,Jung-Wang-Havlin-Kaizoji-Moon-Stanley-2008-EPJB,Wang-Yamasaki-Havlin-Stanley-2008-PRE,Qiu-Guo-Chen-2008-PA,Ren-Zhou-2008-EPL,Ren-Guo-Zhou-2009-PA,Ren-Gu-Zhou-2009-PA}.

Meanwhile, the recurrence interval analysis
is also carried out for stock returns \cite{Yamasaki-Muchnik-Havlin-Bunde-Stanley-2006-inPFE,Bogachev-Eichner-Bunde-2007-PRL,Bogachev-Bunde-2008-PRE,Bogachev-Bunde-2009-PRE,Ren-Zhou-2010-NJP}.
Numerical simulation studies have shown that the long-term correlation of the time
series has a remarkable influence on the recurrence interval
distribution: for the linear long-term correlated time series the recurrence interval
distribution follows a stretched exponential
\cite{Bunde-Eichner-Havlin-Kantelhardt-2003-PA,Bunde-Eichner-Kantelhardt-Havlin-2005-PRL,Altmann-Kantz-2005-PRE,Olla-2007-PRE},
and for the artificial multifractal signals which have non-linear correlation the recurrence interval
distribution exhibits a power-law behavior whose exponent varies with the threshold
\cite{Bogachev-Eichner-Bunde-2007-PRL,Bogachev-Eichner-Bunde-2008-EPJST}. The financial
time series have been reported to have multifractal nature, and the power-law distribution is indeed
observed in empirical studies \cite{Kaizoji-Kaizoji-2004a-PA,Bogachev-Bunde-2008-PRE}. Very recently, Ren and Zhou studied the recurrence intervals $\tau$ between large price returns using the latest data of Chinese stock markets,
and also found that the tail of the recurrence interval distributions could be approximated by a power law \cite{Ren-Zhou-2010-NJP}.

The trading volume is known as an important variable reflecting the liquidity of the financial markets,
and therefore is considered to be of great importance for the measurement of market liquidity risk.
There have been numerous studies conducted to study the relationship
between the trading volumes and the price returns, and growing evidence
shows that large price movements might be driven by large trading
volumes, described well by the price impact function \cite{Hasbrouck-1991-JF,Gopikrishnan-Plerou-Gabaix-Stanley-2000-PRE,Gabaix-Gopikrishnan-Plerou-Stanley-2003-Nature,Lillo-Farmer-Mantegna-2003-Nature,Lim-Coggins-2005-QF,Zhou-2007-XXX}. Some of these studies
have revealed that the trading volumes and the magnitude of the
price returns have universal properties, such as the fat-tailed
distribution and the long-term memory effect \cite{Gopikrishnan-Plerou-Gabaix-Stanley-2000-PRE,Gabaix-Gopikrishnan-Plerou-Stanley-2003-Nature}.
The recurrence interval analysis method has been presently applied to many other fields, e.g., turbulence \cite{Liu-Jiang-Ren-Zhou-2009-PRE} and networks \cite{Bogachev-Bunde-2009-EPL,Cai-Fu-Zhou-Gu-Zhou-2009-EPL}. In this paper, we attempt to study the recurrence intervals between large trading volumes, and test if recurrence intervals of the trading volumes show power-law distribution and memory effects similar to those of the price returns.

The paper is organized as follows. In Section 2, we introduce the
data sets analyzed and the investigated variables. Sections 3 and 4
study the probability distribution and the memory effects of
the recurrence intervals respectively. In Section 5, we study the
relationship between trading volumes and price returns based on the recurrence
interval analysis. Section 6 gives the conclusion.

\section{Data sets}

We analyze the 1-min intraday data of $20$ liquid stocks actively traded
on the Shanghai Stock Exchange and the Shenzhen Stock Exchange from
January 2000 to May 2009, the Shanghai Stock Exchange Composite
Index (SSEC) and the Shenzhen Stock Exchange Composite Index (SZCI)
from January 2003 to April 2009. The data are retrieved from a database developed by GTA Information Technology Co., Ltd, see http://www.gtadata.com/.
Since the sampling time is 1 minute, the total number of data points is about 340000 for the two Chinese
indices and 500000 for individual stocks. These $20$ stocks are
actively traded stocks representative in a variety of industry
sectors. Each stock is uniquely identified with a stock code which
is a unique 6-digit number. A stock with the code initiated with
$60$ is traded on the Shanghai Stock Exchange, while a stock with
the code initiated with $00$ is listed on the Shenzhen Stock
Exchange.

\begin{figure}[htb]
\centering
\includegraphics[width=8cm]{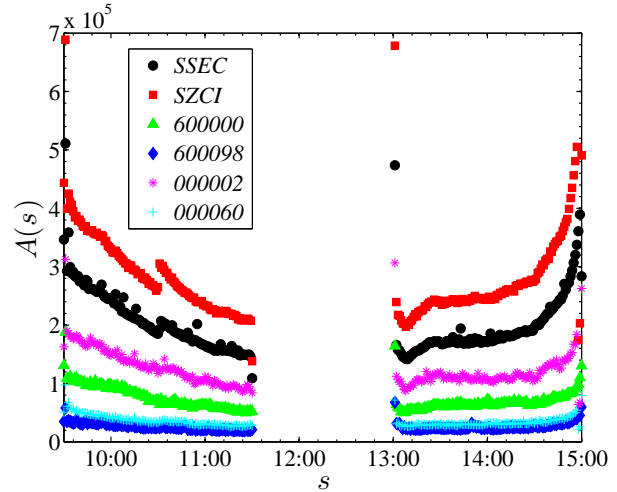}
\caption{\label{Fig:IntradayPattern}(Color online) Interval intraday patterns of trading volumes for SSEC, SZCI and four representative
stocks. }
\end{figure}

The 1-min trading volumes of the Chinese stock markets exhibit a
U-shaped intraday pattern, like many Western stock markets \cite{Wang-Yamasaki-Havlin-Stanley-2006-PRE,Wang-Weber-Yamasaki-Havlin-Stanley-2007-EPJB,VodenskaChitkushev-Wang-Weber-Yamasaki-Havlin-Stanley-2008-EPJB,Wang-Yamasaki-Havlin-Stanley-2008-PRE}.
The intraday pattern is defined as
\begin{equation}
A(s)=\frac{\sum_{i=1}^{N} V_i(s)}{N}, \label{e10}
\end{equation}
which is the average volume at a specific minute $s$ of the trading
day averaged over all $N$ trading days and $V_i(s)$ is the trading
volume at time $s$ of day $i$. The intraday patterns of the 1-min
trading volumes for the two Chinese indices and four representative stocks
are illustrated in Fig.~\ref{Fig:IntradayPattern}. The Shanghai Stock
Exchange and Shenzhen Stock Exchange open at 9:30 a.m., and close at
15:00 p.m., during which there exists a midday break between 11:30
a.m. and 13:00 p.m.. Similar to many Western stock markets, the average
volume of the Chinese stock markets displays sharp peaks close to the
opening and closing times. In addition, the average volume of the
Chinese stock market surges soon after midday break which may due to
the information aggregation and the cumulative orders submitted during the midday break. Taking a more
careful look at the curves for the two Chinese indices, one observes
that the average volume shows a relatively sharp jump at around
10:30 a.m.. This phenomenon can be explained by the trading rule that
for the stocks which behaved abnormally the Stock Exchange will
suspend trading in relevant securities. Until the day
the parties with disclosure obligation make relevant
announcements, the stocks involved will resume trading at 10:30 a.m. on that day. All these factors will affect
the measurement of the trading volumes, and will consequently lead to periodic
behavior in the distribution of recurrence intervals between large
trading volumes.

To avoid the effect of this periodic oscillation, the intraday
pattern is removed by
\begin{equation}
V'(t)=\frac{V(t)}{A(s)}. \label{e20}
\end{equation}
Then the volatility is normalized by dividing its standard deviation
\begin{equation}
v(t)=\frac{V'(t)}{[\langle V'(t)^2 \rangle-\langle V'(t)
\rangle^2]^{1/2}}. \label{e30}
\end{equation}
In this paper, we mainly focus on the statistic properties of the
recurrence intervals between normalized trading volumes exceeding
a threshold $q$. For a certain value of $q$, a series of
recurrence intervals $\tau_q$ are obtained, therefore we can
study the statistic properties of the
recurrence intervals for different values of $q$.

\section{Probability distribution of recurrence intervals between large trading volumes}

\subsection{Empirical distribution}

It is well accepted that there exists a close relationship between price returns and trading volumes.
Recent empirical studies on the recurrence intervals between price returns have shown
that the tail of the recurrence interval distribution follows a
power-law scaling. Here we study the probability distribution
function (PDF) $P_q(\tau)$ of the recurrence intervals between
trading volumes, and see if the power-law scaling behavior
maintains.

\begin{figure}[htb]
\centering
\includegraphics[width=4cm]{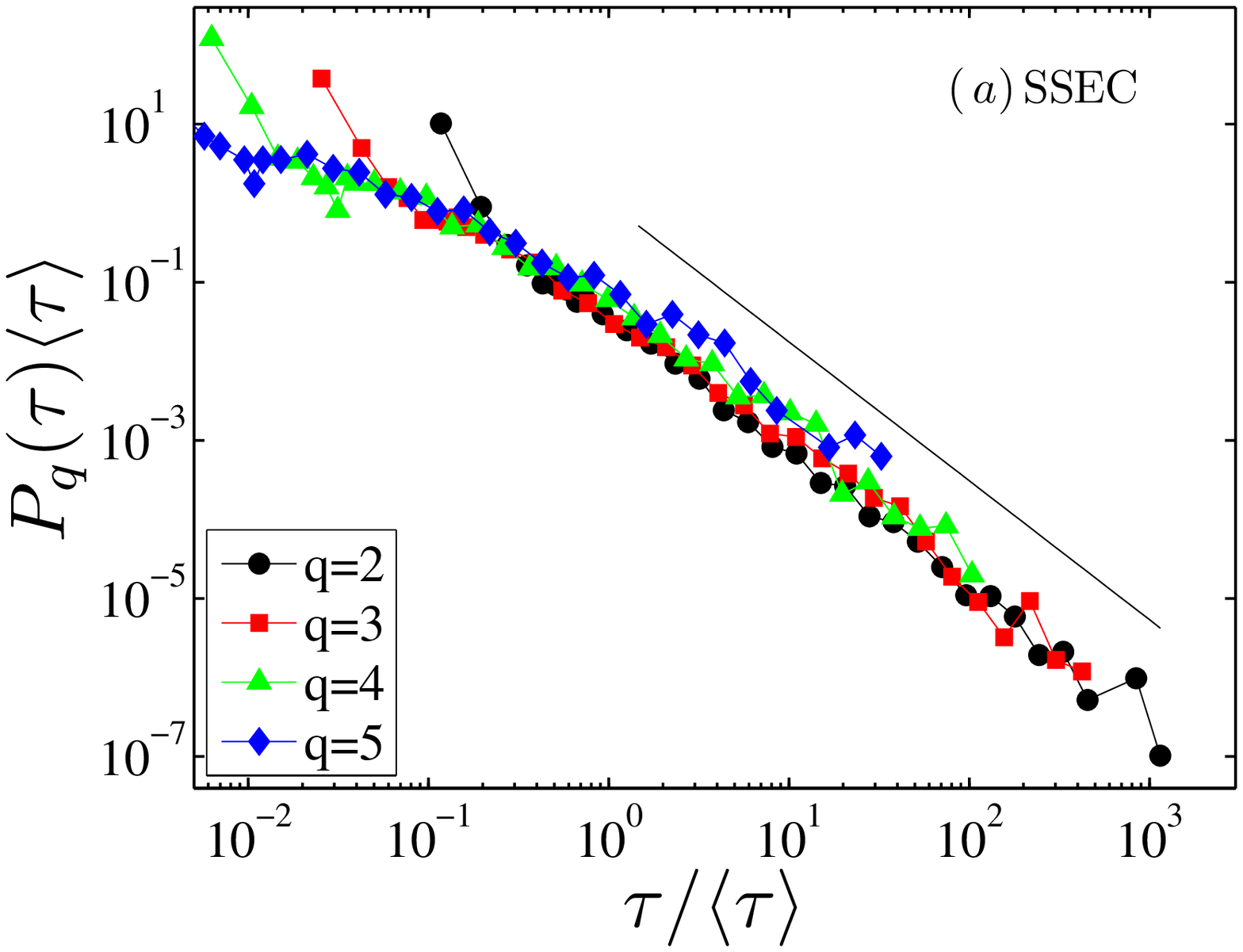}
\includegraphics[width=4cm]{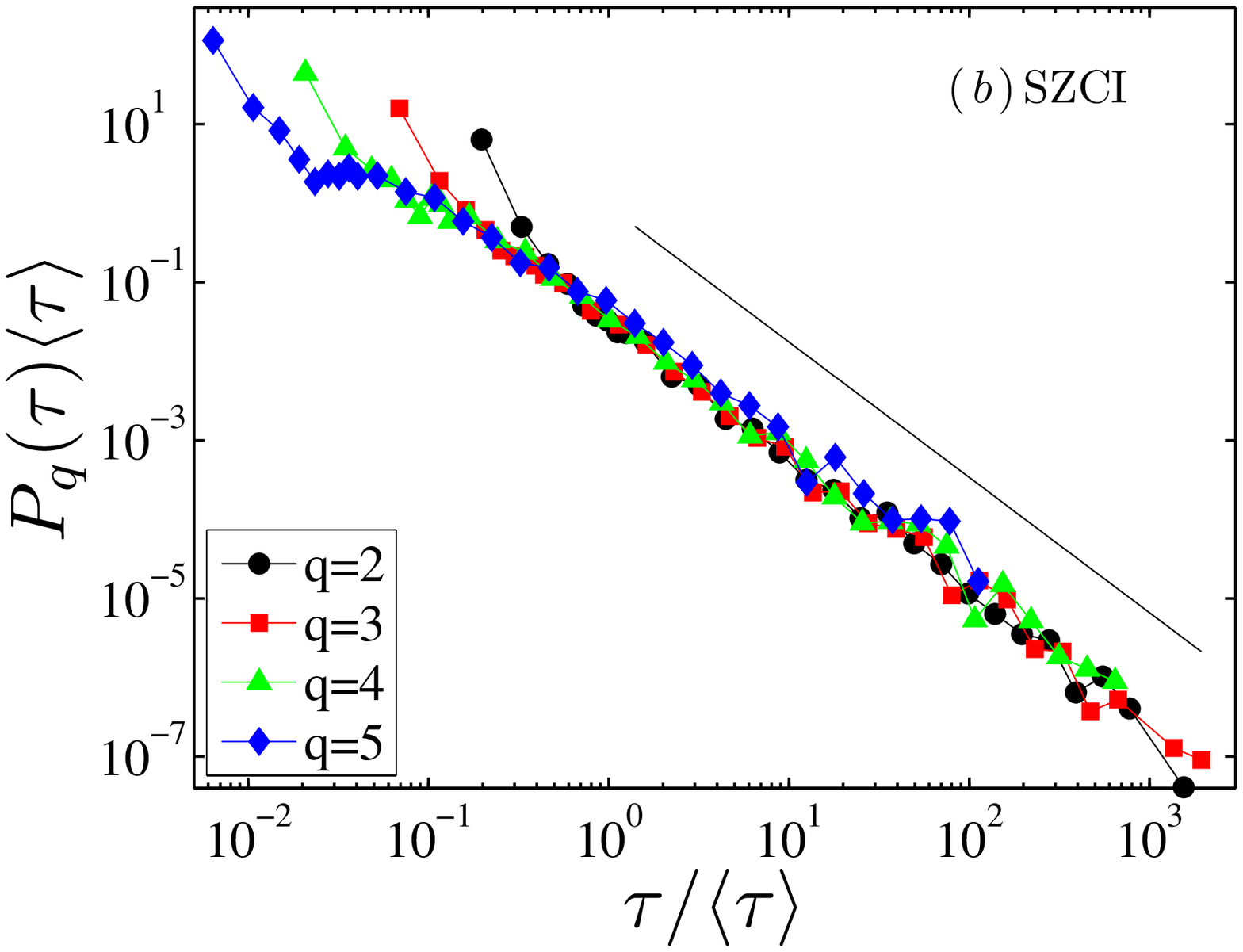}
\includegraphics[width=4cm]{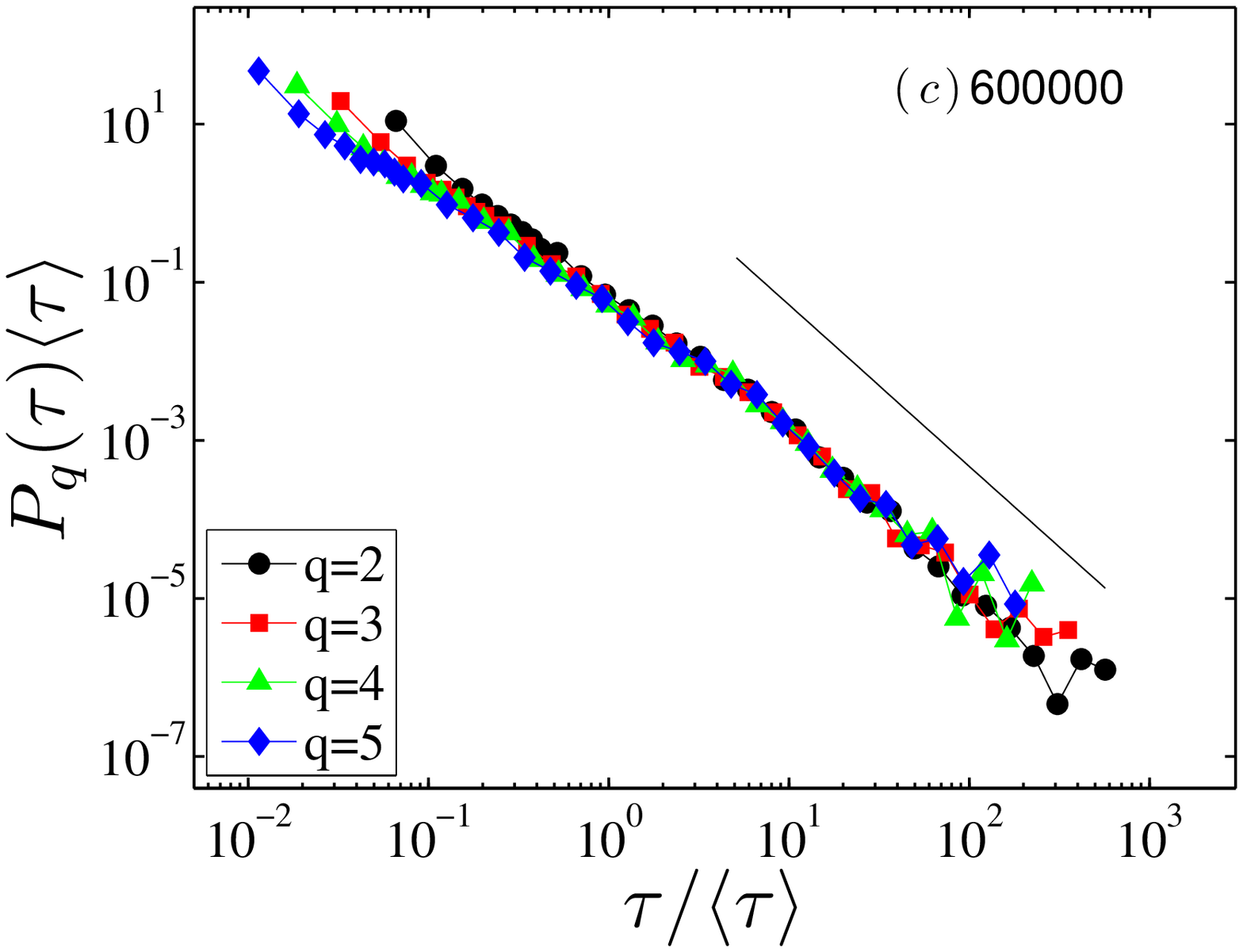}
\includegraphics[width=4cm]{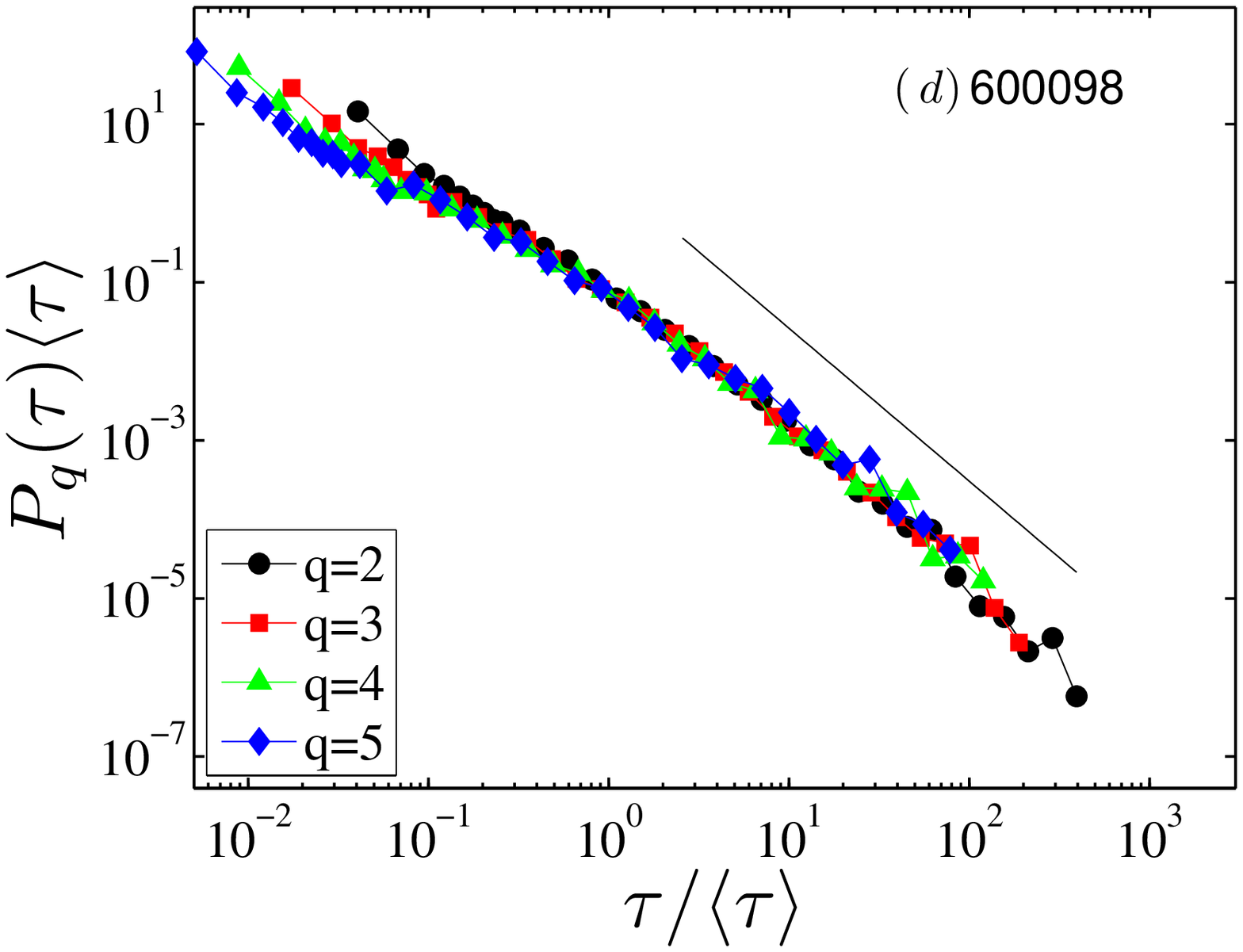}
\includegraphics[width=4cm]{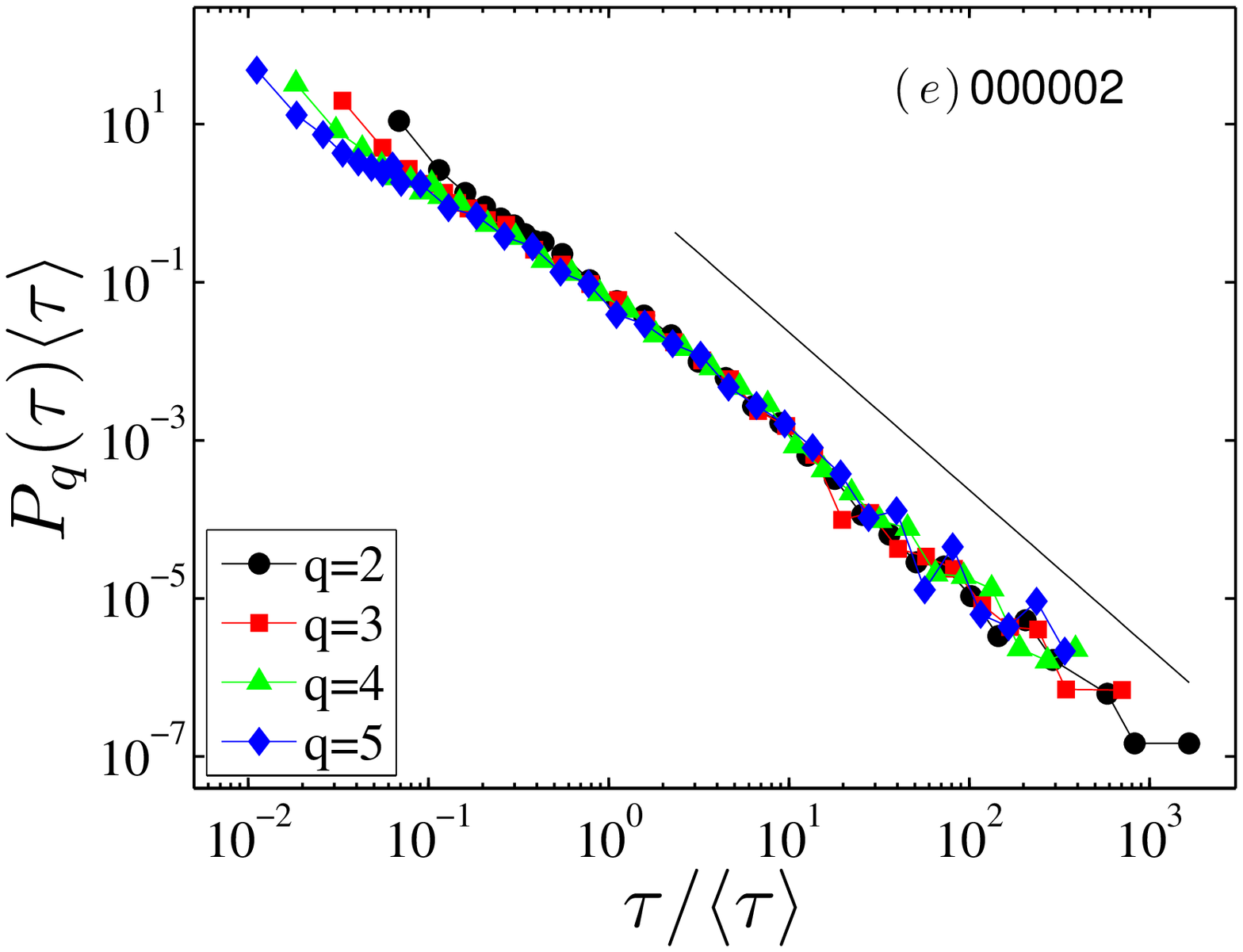}
\includegraphics[width=4cm]{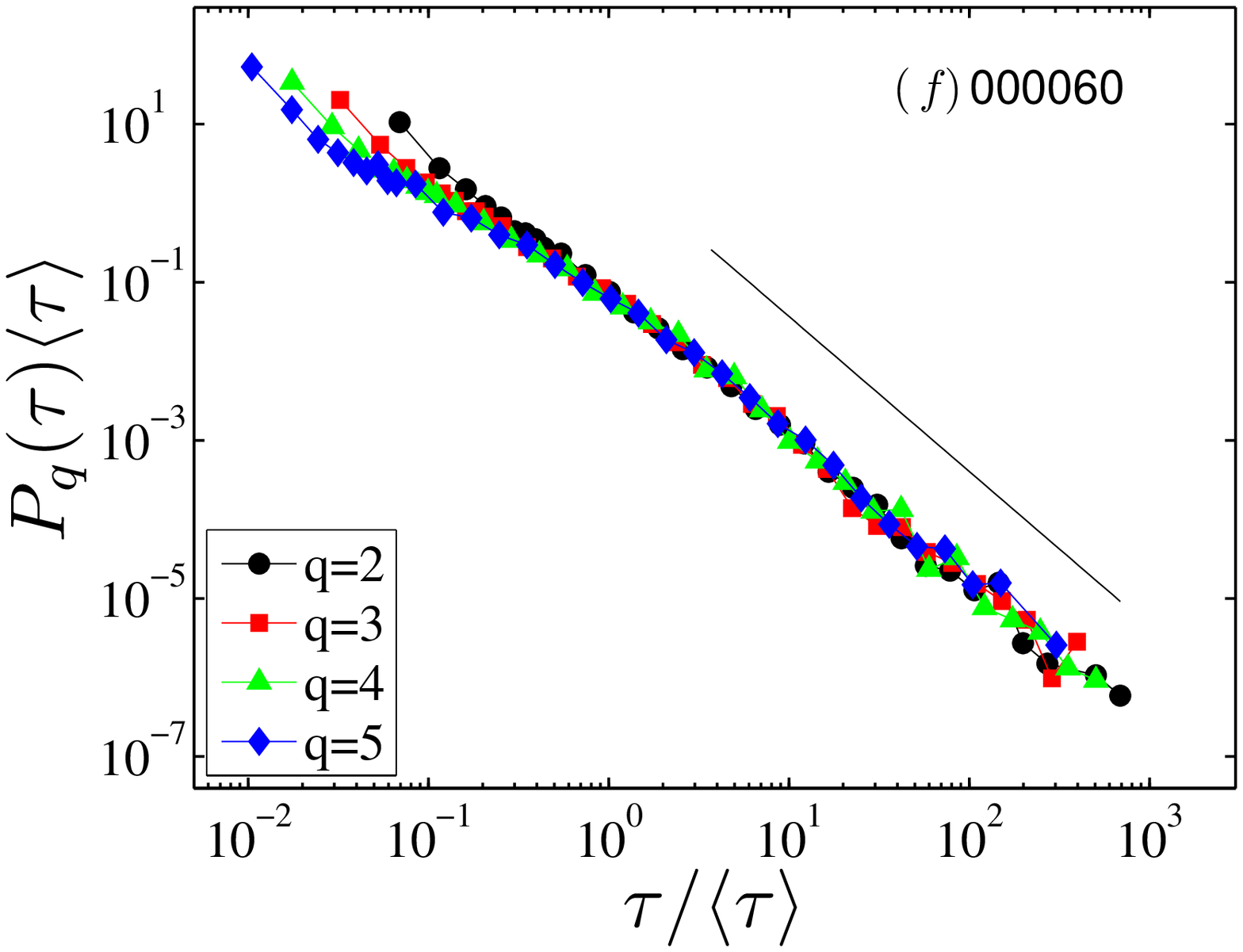}
\caption{\label{Fig:RI:PDF} (Color online) Empirical probability
distributions of scaled recurrence intervals for different thresholds
$q=2,3,4,5$ for SSEC, SZCI and four representative stocks. The solid
curves are the fitted functions $c x^{-\delta}$ with parameters
listed in Table~\ref{TB:goodness-of-fit-test}.}
\end{figure}

In Fig.~\ref{Fig:RI:PDF}, the scaled PDFs $P_q(\tau) \langle \tau
\rangle$ are plotted as a function of the scaled recurrence interval $\tau/
\langle \tau \rangle$ for various values of $q=2,3,4,5$ for the two
Chinese indices and four representative stocks. For small
values of $\tau/ \langle \tau \rangle$, $P_q(\tau) \langle \tau
\rangle$ diverges for different $q$ values, which may partially due
to the discreteness of recurrence intervals. However, one can also observe that for large values of $\tau/ \langle \tau
\rangle$ the curves for different $q$ values almost collapse onto a
single curve and show nice linear behavior in the double logarithmic
plot. Hence, we can conclude that the tails of the PDFs of the
recurrence intervals between trading volumes follow a scaling
behavior as
\begin{equation}
P_q(\tau)= \frac{1} {\langle \tau \rangle} f ( \tau/\langle \tau
\rangle ). \label{Eq:Pq:f}
\end{equation}
This indicates that the scaled PDF for large values of the scaled
interval is independent of the threshold $q$.

\subsection{Determining the scaling function}

The linear behavior of the scaled PDF $P_q(\tau) \langle \tau
\rangle$ for large values of $\tau/ \langle \tau \rangle$ in the
double logarithmic plot suggests that the scaling function may
follow a power law as
\begin{equation}
  f(\tau/\langle \tau \rangle)=f(x)=c x^{-\delta},\ x\geq x_{\min}
  \label{Eq:PL}
\end{equation}
for $x$ larger than a lower bound $x_{\min}$. We aggregate the
interval samples above $x_{\min}$ for different values of
$q=2,3,4,5$, and fit them using a common power-law function presented in
Eq.~(\ref{Eq:PL}).

Recently, an efficient method of fitting power-law distributions
based on the Kolmogorov-Smirnov (KS) statistic is proposed by Clauset,
Shalizi and Newman \cite{Clauset-Shalizi-Newman-2009-SIAMR}. We use
this method with the purpose of making a relatively accurate
estimation of the parameters in Eq.~(\ref{Eq:PL}). The $KS$
statistic is defined as
\begin{equation}
   KS = \max_{x\geq \hat{x}_{\min}} \left(|F-F_{\rm{PL}}|\right),
   \label{Eq:KS}
\end{equation}
where $F$ is the cumulative distribution function (CDF) of the
empirical data and $F_{\rm{PL}}$ is the CDF of the power-law fit. To
make the empirical PDF and the best power-law fit as similar as
possible, we determine the estimate of $\hat{x}_{\min}$ by minimizing
the $KS$ statistic, then we estimate the values of the other two
parameters $c$ and $\delta$ using the maximum likelihood method. The
estimated parameters $\hat{x}_{\min}$, $\delta$, $c$ and the resultant
$KS$ statistic are depicted in Table~\ref{TB:goodness-of-fit-test},
and the power-law fits are correspondingly illustrated in
Fig.~\ref{Fig:RI:PDF}. For the two indices, $\hat{x}_{\min}$
shows small values close to $1.4$ indicating a scaling region over almost three orders
of magnitude with power-law exponents $\delta = 1.74\pm0.04$. Among $20$
stocks, $16$ stocks have $\hat{x}_{\min}<9.0$ indicating a scaling region over more than two
orders of magnitude with power-law exponents $\delta = 2.0\pm0.3$.
The rest of stocks (600009, 600058, 600104 and 000063) have $\hat{x}_{\min} > 9.0$,
and show scaling regions spanning almost two
orders of magnitude with exponents larger than $2.0$.

\begin{table}[htp]
 \centering
 \caption{Estimated parameters for the power-law fit of empirical PDFs, and goodness-of-fit tests
based on $KS$, $KSW$ and $CvM$ statistics for two Chinese indices and $20$ stocks. }
\label{TB:goodness-of-fit-test}
\begin{tabular}{lccccccc}
  \hline
   Code & $\hat{x}_{\min}$ & $\delta$ & $c$ & $KS$ & $p_{KS}$ & $p_{KSW}$ & $W^2$ \\
  \hline
   SSEC &  $1.46$ & $1.76(0.02)$ & $0.05$ & $0.0257$ & $0.121$ & $0.016$ & $0.17$\\%
   SZCI$^\dagger$&  $1.40$ & $1.71(0.01)$ & $0.03$ & $0.0285$ & $0.038$ & $\bf0.003$ & $0.26$\\%
 $600000$& $5.12$ & $2.04(0.03)$ & $0.16$ & $0.0143$ & $0.877$ & $0.290$ & $0.06$\\%
 $600009$ & $20.27$ & $2.75(0.09)$ & $2.50$ & $0.0328$ & $0.635$ & $0.750$ & $0.07$\\%
 $600016$ & $1.82$ & $2.08(0.02)$ & $0.12$ & $0.0222$ & $0.032$ & $0.010$ & $0.25$\\%
 $600058$ &$15.06$ & $2.17(0.06)$ & $0.26$ & $0.0196$ & $0.982$ & $0.999$ & $0.02$\\%
 $600060^\star$ & $1.85$ & $1.92(0.02)$ & $0.13$ & $0.0347$ &    $\bf0$ &  $\bf0.004$ & $\bf1.65$\\%
 $600073$ & $4.53$ & $2.03(0.03)$ & $0.17$ & $0.0191$ & $0.650$ & $0.158$ & $0.09$\\%
 $600088^\star$ & $2.95$ & $1.99(0.02)$ & $0.16$ & $0.0381$ &    $\bf0$ & $\bf0.008$ & $\bf1.42$\\%
 $600098$ & $2.57$ & $1.93(0.02)$ & $0.12$ & $0.0312$ & $0.031$ & $0.055$ & $0.35$\\%
 $600100$ & $7.54$ & $1.97(0.03)$  &$0.14$ & $0.0316$ & $0.176$ & $0.145$ & $0.26$\\%
 $600104$ & $9.24$ & $2.59(0.05)$ & $1.15$ & $0.0266$ & $0.191$ & $0.159$ & $0.17$\\%
 $000001^\dagger$ & $2.73$ & $1.94(0.02)$ & $0.13$ & $0.0311$ & $\bf0.007$ & $0.029$ & $0.72$\\%
 $000002^\dagger$ & $2.33$ & $2.00(0.02)$ & $0.11$ & $0.0197$ & $0.171$ &     $\bf0$ & $0.27$\\%
 $000021^\dagger$ & $0.72$ & $1.75(0.01)$ & $0.06$ & $0.0215$ & $0.019$ &     $\bf0$ & $0.61$\\%
 $000027^\star$ & $0.78$ & $1.79(0.01)$ & $0.09$ & $0.0348$ &    $\bf0$ &    $\bf0$ & $\bf2.93$\\%
 $000029$ & $6.38$ & $2.12(0.04)$ & $0.24$ & $0.0229$ & $0.508$ & $0.220$ & $0.10$\\%
 $000031$ & $2.31$ & $1.89(0.02)$ & $0.09$ & $0.0202$ & $0.320$ & $0.064$ & $0.16$\\%
 $000039$ & $2.37$ & $2.03(0.02)$ & $0.12$ & $0.0133$ & $0.733$ & $0.111$ & $0.07$\\%
 $000060$ & $3.71$ & $1.96(0.03)$ & $0.11$ & $0.0146$ & $0.816$ & $0.274$ & $0.03$\\%
 $000063$ & $37.88$ & $3.27(0.19)$ &$23.68$ & $0.0258$ & $0.997$ & $0.311$ & $0.01$\\%
 $000066^\star$ & $0.56$ & $1.72(0.01)$ & $0.07$ & $0.0305$ &    $\bf0$ &    $\bf0$ & $\bf1.57$\\%
  \hline
\end{tabular}
\end{table}

We have performed a power-law fit for the empirical interval
distributions, and given a relatively accurate estimation of the
parameters. Furthermore, a goodness-of-fit test should be adopted to
examine the goodness of the power-law fit. Based upon the $KS$
statistic, we test the
hypothesis that the empirical PDFs could be fitted well by their
common power-law fit. To do this, a bootstrap method is adopted
following Refs.~
\cite{Clauset-Shalizi-Newman-2009-SIAMR,Gonzalez-Hidalgo-Barabasi-2008-Nature}.
One thousand synthetic data sets are randomly generated from the best
power-law fit of the empirical PDFs. For each synthetic data set, a
$KS$ statistic is obtained as
\begin{equation}
   KS_{\rm{sim}} = \max\left(|F_{\rm{sim}}-F_{\rm{sim,PL}}|\right),
   \label{Eq:KS:sim}
\end{equation}
where $F_{\rm{sim}}$ is the CDF
of the synthetic data and $F_{\rm{sim,PL}}$ is the CDF of the power-law
fit for synthetic data. The $KS$ statistic for the empirical data is
defined by Eq.~(\ref{Eq:KS}). The $p$-value is defined as the
frequency that $KS_{\rm{sim}}>KS$, which is regarded as the
probability that the power-law fit is coincident with the empirical
PDF.

A variant of the $KS$ statistic, known as the $KSW$ statistic
\cite{Gonzalez-Hidalgo-Barabasi-2008-Nature}, is also used to
perform the goodness-of-fit test. The $KSW$ statistic for the
empirical data is defined as
\begin{equation}
KSW = \max_{x\geq \hat{x}_{\min}}
 \left( \frac{|F-F_{\rm{PL}}|} {\sqrt{F_{\rm{PL}}(1-F_{\rm{PL}})}} \right). \label{Eq:KSW}
\end{equation}
The $KSW$ statistic with this definition is more sensitive on the
edges of the cumulative distributions. In the bootstrap process, 1000
synthetic data sets are generated from the best power-law fit,
and therefore the $KSW$ statistic for the synthetic data is calculated
by
\begin{equation}
KSW_{\rm{sim}} = \max
 \left( \frac{|F_{\rm{sim}}-F_{\rm{sim,PL}}|} {\sqrt{F_{\rm{sim,PL}}(1-F_{\rm{sim,PL}})}} \right). \label{Eq:KSW:sim}
\end{equation}
Similar to the $p$-value for $KS$ statistic, the $p$-value for the
$KSW$ statistic could be obtained by calculating the frequency that
$KSW_{\rm{sim}}>KSW$.

The goodness-of-fit tests based on the $KS$ and $KSW$ statistics are
carried out for the two Chinese indices and $20$ individual stocks,
and the resultant $p$-values are depicted in
Table~\ref{TB:goodness-of-fit-test}. Consider the significance level
of 1\%, if the $p$-value is larger than 0.01, we can conclude that
the stock passes the test, and consequently the null hypothesis that
the empirical PDFs of the relevant stock for different $q$ values
could be fitted well by their common power-law fit is accepted.
Among the $20$ stocks, four stocks (600060,600088,000027,000066) marked
with $\star$ fall in the test using both $KS$ and $KSW$ statistics,
and three stocks (000001,000002,000021) marked with $\dagger$ fail in
the test using $KS$ or $KSW$ statistic. In other words, 13 stocks
pass the test using both $KS$ and $KSW$ statistics, and 16 stocks
pass the test using at least one of $KS$ and $KSW$ statistics. Based
upon this fact, we may roughly conclude that for most of the stocks
the tails of empirical PDFs follow scaling behavior and the
scaling function could be approximated by a power law. Similar
results are obtained for the two Chinese indices composed of
individual stocks traded on relevant stock exchanges: SSEC passes the
goodness-of-fit test using both $KS$ and $KSW$ statistics, and SZCI
passes the goodness-of-fit test using $KS$ statistic.

We also use other variation of the goodness-of-fit test on place of
the $KS$ statistic. The Cram\'{e}r-von Mises (CvM) statistic is
another common used statistic, which is defined as
\begin{equation}
   W^2 = N \int_{-\infty}^{\infty} ( F-F_{\rm{PL}})^2 d F_{\rm{PL}},
   \label{Eq:CvM}
\end{equation}
where $F$ and $F_{PL}$ are the CDFs of the empirical data and its
power-law fit respectively, and $N$ is the number of scaled interval
samples $x=\tau / \langle \tau \rangle$
\cite{Pearson-Stephens-1962-Bm,Stephens-1964-Bm,Stephens-1970-JRSSB}.
For a sequence of scaled interval samples  $x_1, x_2,\cdots,x_N$
arranged in ascending order, the computational form of $W^2$
statistic is given by
\begin{equation}
   W^2 = \frac{1}{12N} + \sum_{i=1}^{N} \left( x_i - \frac{2i-1}{2N} \right)^2.
   \label{Eq:CvM:computation}
\end{equation}
In Table~\ref{TB:goodness-of-fit-test}, $W^2$ for the two Chinese
indices and $20$ individual stocks is also depicted. For a stock
which has a $W^2$ smaller than the critical value $0.743$, it
successfully passes the goodness-of-fit test under the significance
level of 1\%. From the table, only four stocks
(600060,600088,000027,000066) show $W^2$ larger than the critical
value, thus fail in the CvM test. In general, the CvM statistic
offers results very similar to those of the $KS$ statistic, and for
the stock which has a large $p$-value close to $1.0$ the $W^2$
statistic of the relevant stock is significantly smaller than the
critical value.

\section{Memory effects in recurrence intervals between trading volumes}
\subsection{Conditional PDF and mean conditional recurrence interval}

Besides the investigation of the probability distribution, the computation of
the temporal correlation offers another important way of understanding the statistical properties of
the recurrence intervals between trading volumes. Previous studies have
detected the presence of memory effects in the recurrence intervals
between price returns \cite{Bogachev-Eichner-Bunde-2007-PRL,Bogachev-Bunde-2008-PRE,Ren-Zhou-2010-NJP}. We thus conjecture similar memory
effects may also exist in the recurrence intervals between trading
volumes.

\begin{figure}[htb]
\centering
\includegraphics[width=4cm]{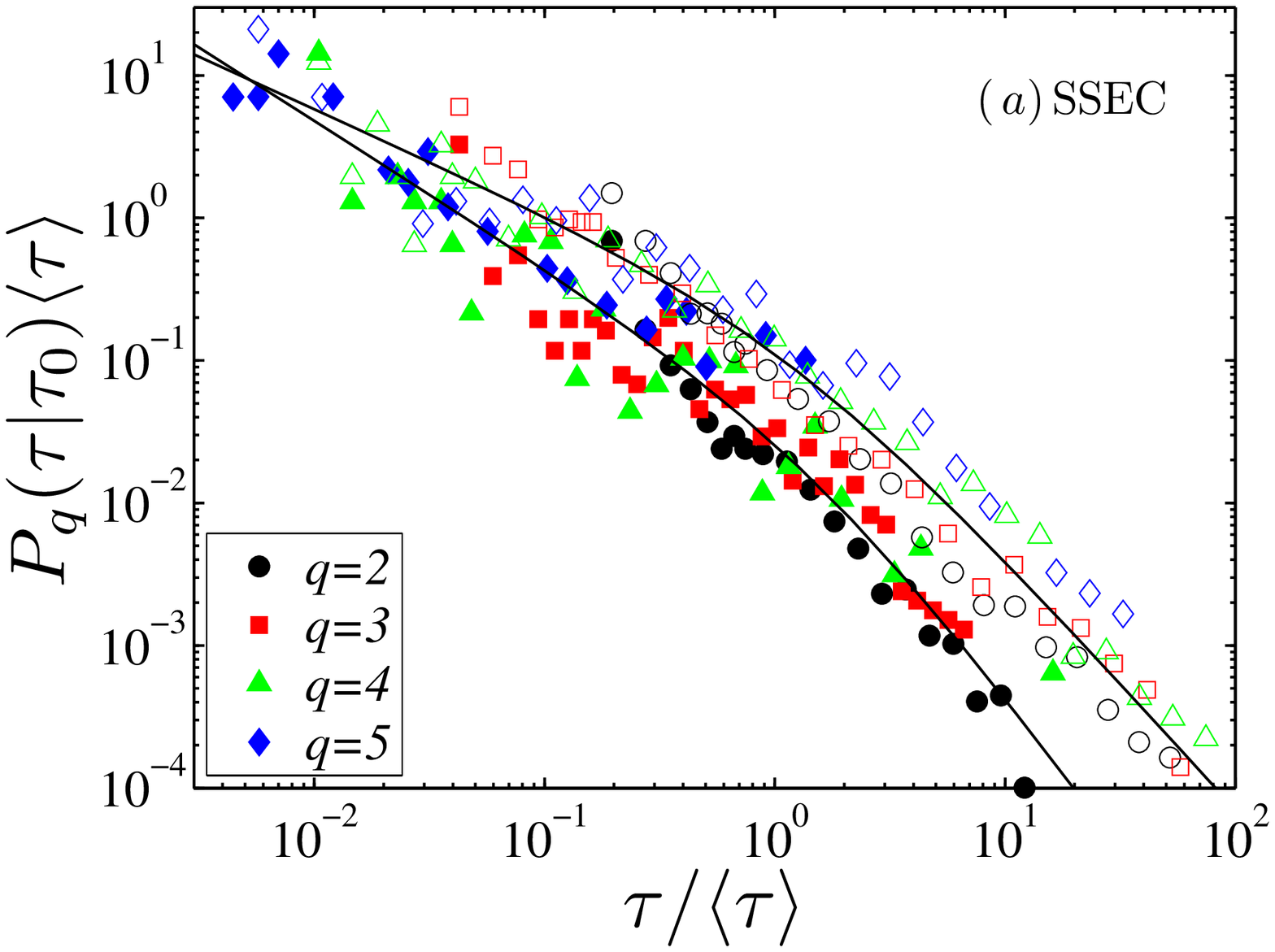}
\includegraphics[width=4cm]{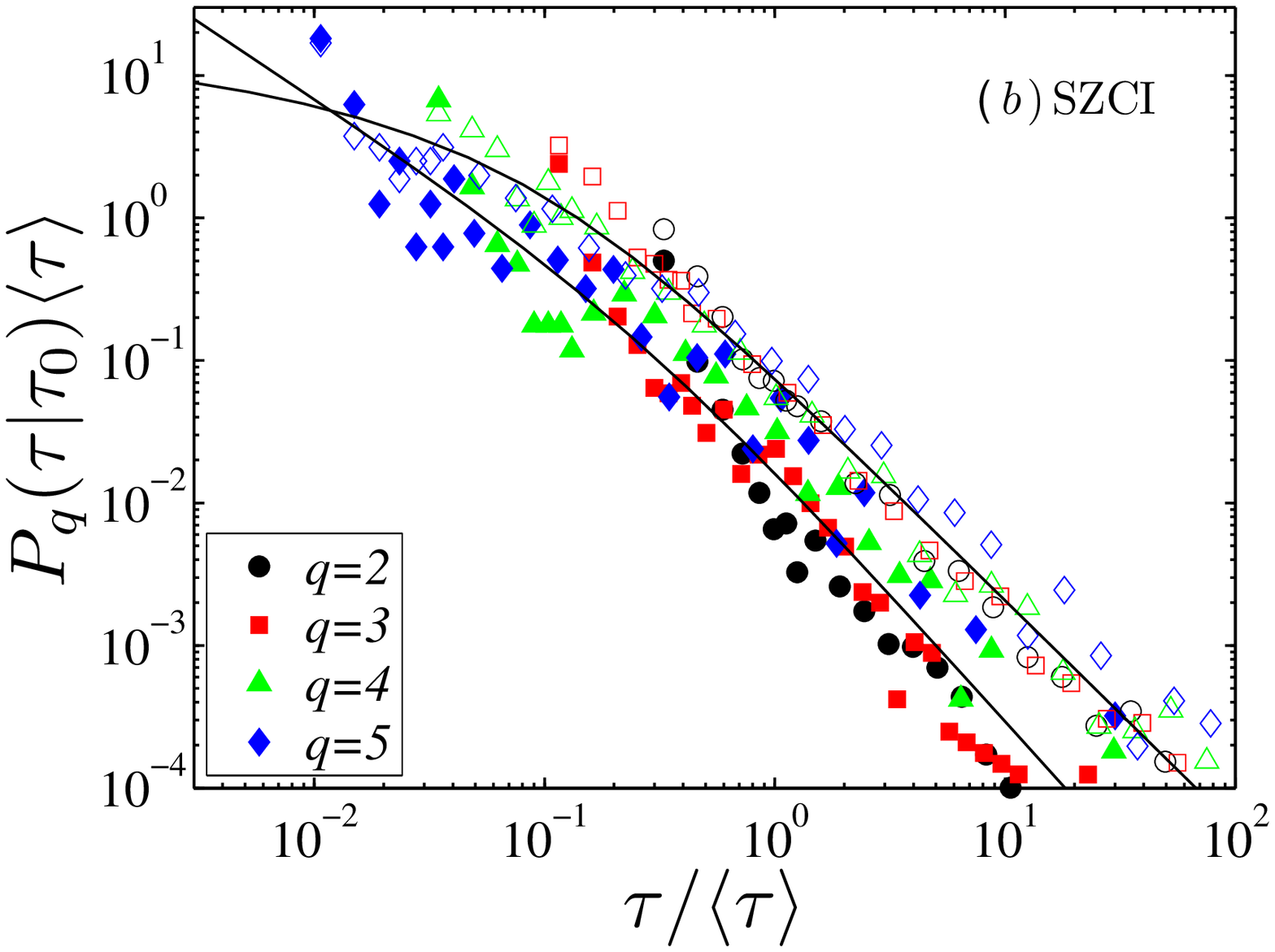}
\includegraphics[width=4cm]{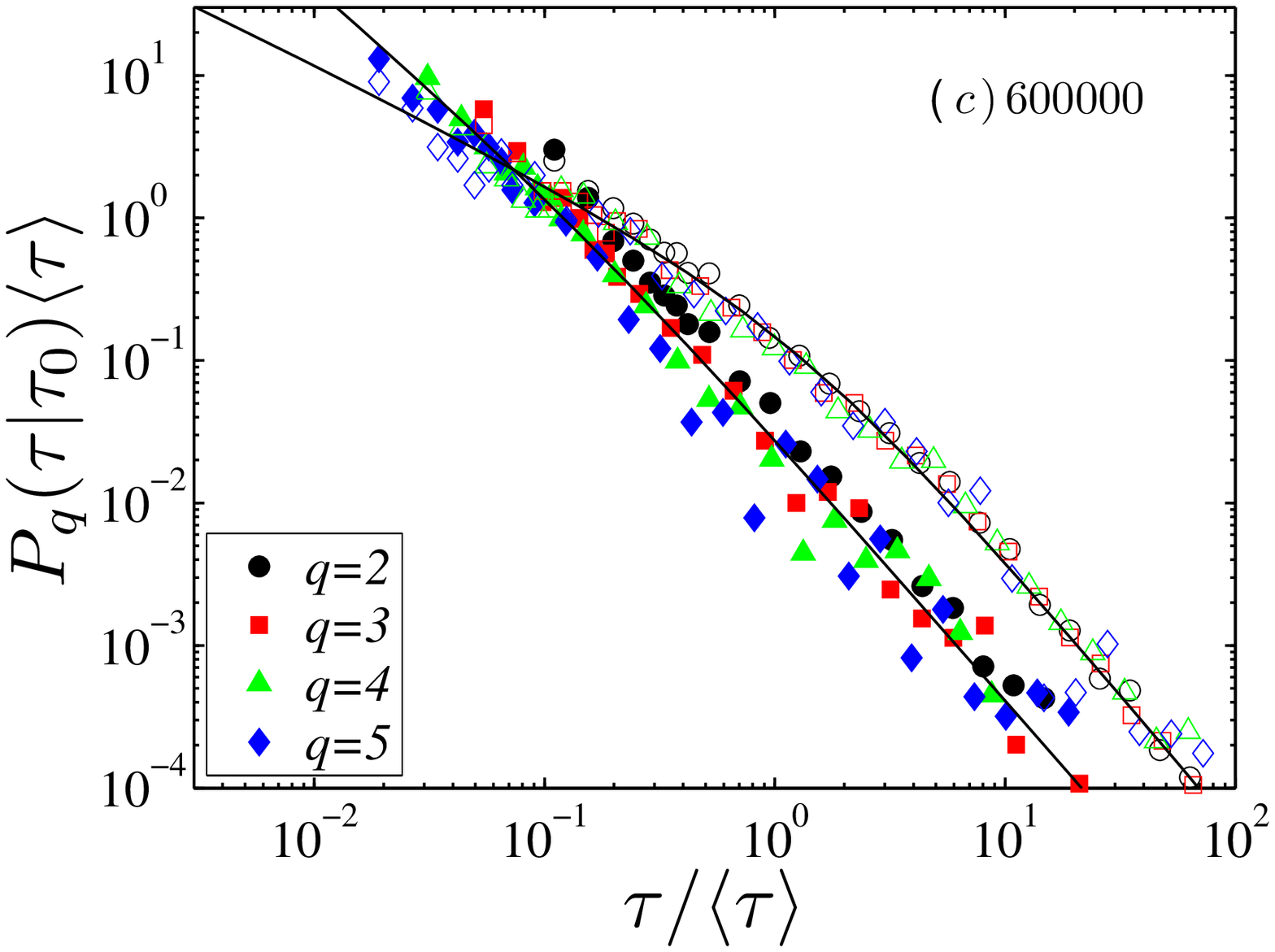}
\includegraphics[width=4cm]{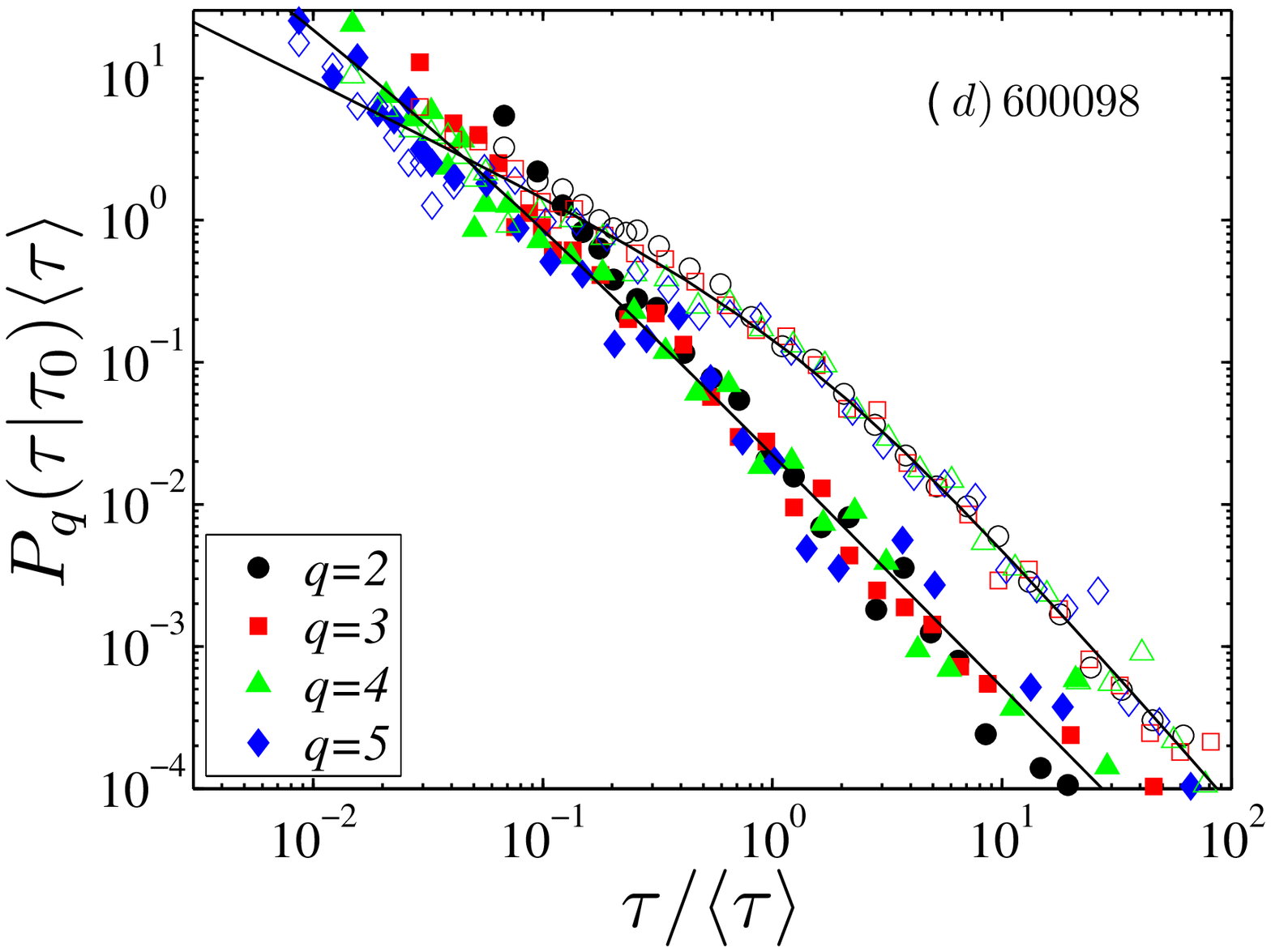}
\includegraphics[width=4cm]{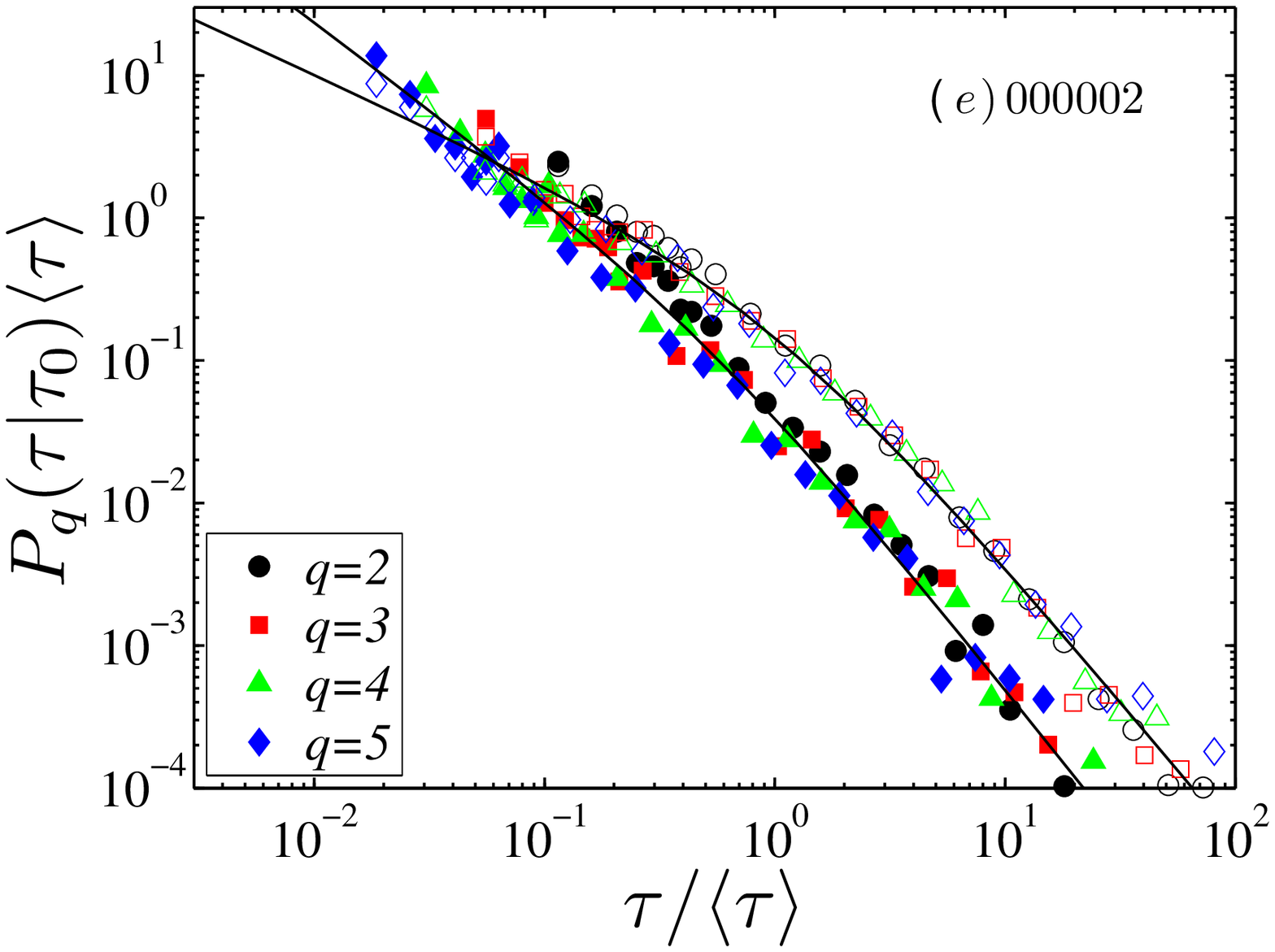}
\includegraphics[width=4cm]{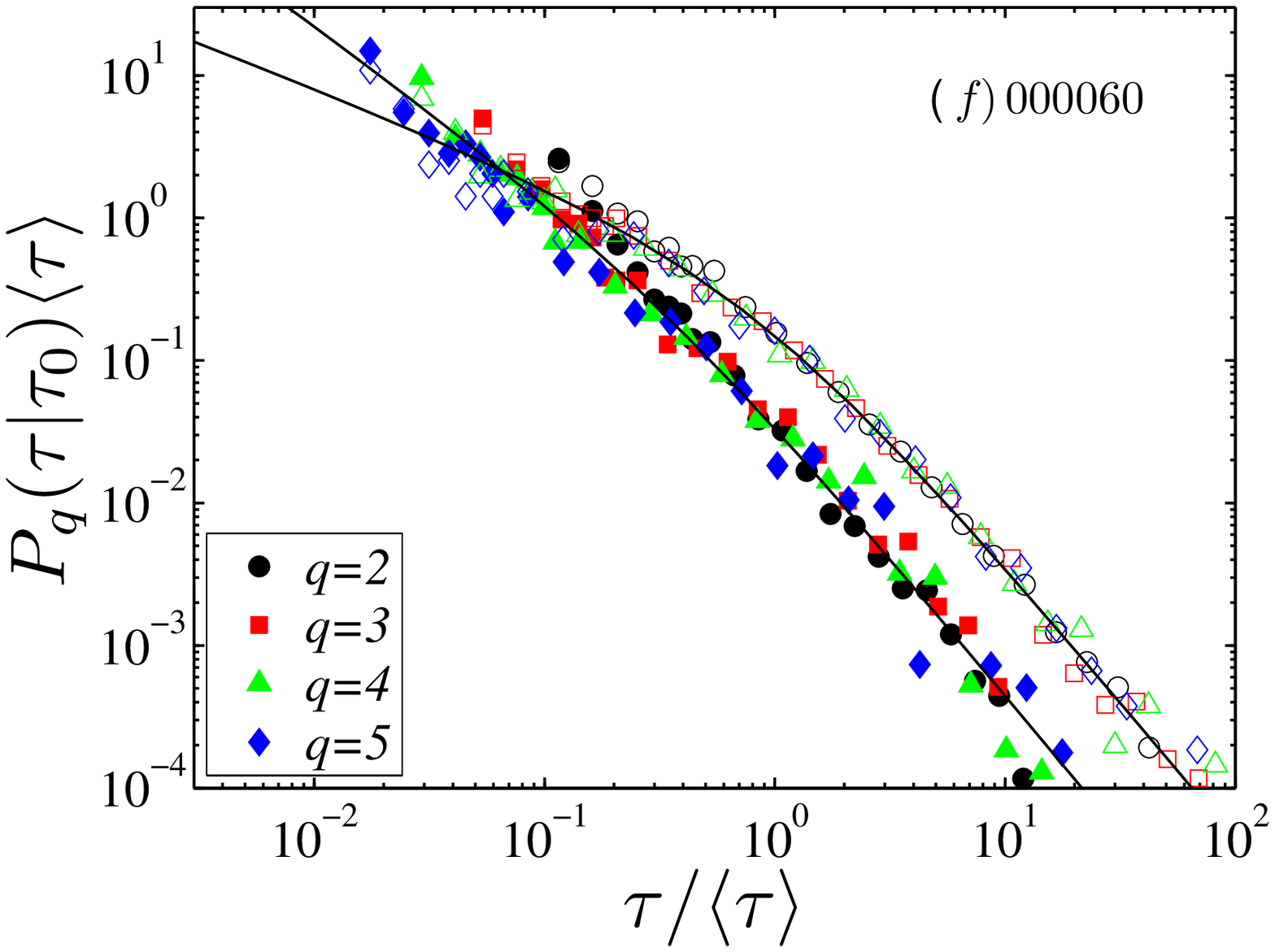}
\caption{\label{Fig:RI:ConPDF} (Color online) Scaled conditional PDF
$P_q(\tau | \tau_0) \langle \tau \rangle$ of scaled return interval
$\tau/\langle \tau \rangle$ with $\tau_0$ in the largest $1/4$
subset (open symbols) and the smallest $1/4$ subset (filled symbols) for
SSEC, SZCI and $4$ representative stocks. The solid lines are
guidelines for looking.}
\end{figure}

To verify our conjecture, we first calculate the conditional PDF of
the recurrence intervals between trading volumes. The conditional
PDF computes the probability of finding a recurrence interval
$\tau$ conditioned on the preceding interval $\tau_0$. The
conditional PDF is calculated for a bin of $\tau_0$ to get better
statistics. We rank the whole sequence of the recurrence intervals
in an ascending order, and partition it to four bins with equal
size. Fig.~\ref{Fig:RI:ConPDF} plots the scaled conditional PDF
$P_q(\tau | \tau_0) \langle \tau \rangle$ as a function of the
scaled recurrence interval $\tau/\langle \tau \rangle$ with
$\tau_0$ in the largest and smallest bins for different $q$-values
for SSEC, SZCI and four representative stocks. The curves for
different $q$-values with $\tau_0$ in the largest and smallest bins
approximately collapse on to two separate solid curves as
illustrated in Fig.~\ref{Fig:RI:ConPDF}. For small $\tau/\langle
\tau \rangle$, $P_q(\tau | \tau_0) \langle \tau \rangle$ with
$\tau_0$ in the smallest bin is larger than that with $\tau_0$ in
the largest bin, while $P_q(\tau | \tau_0) \langle \tau \rangle$ with $\tau_0$ in the
largest bin shows values larger than that with $\tau_0$ in the smallest bin when
$\tau/\langle \tau \rangle$ is large. This means that small $\tau_0$
tends to be followed by small $\tau$ and large $\tau_0$ tends to be
followed by large $\tau$, and therefore provides an evidence of
the short-term memory.

\label{S2:CondPDF}

\begin{figure}[htb]
\centering
\includegraphics[width=4cm]{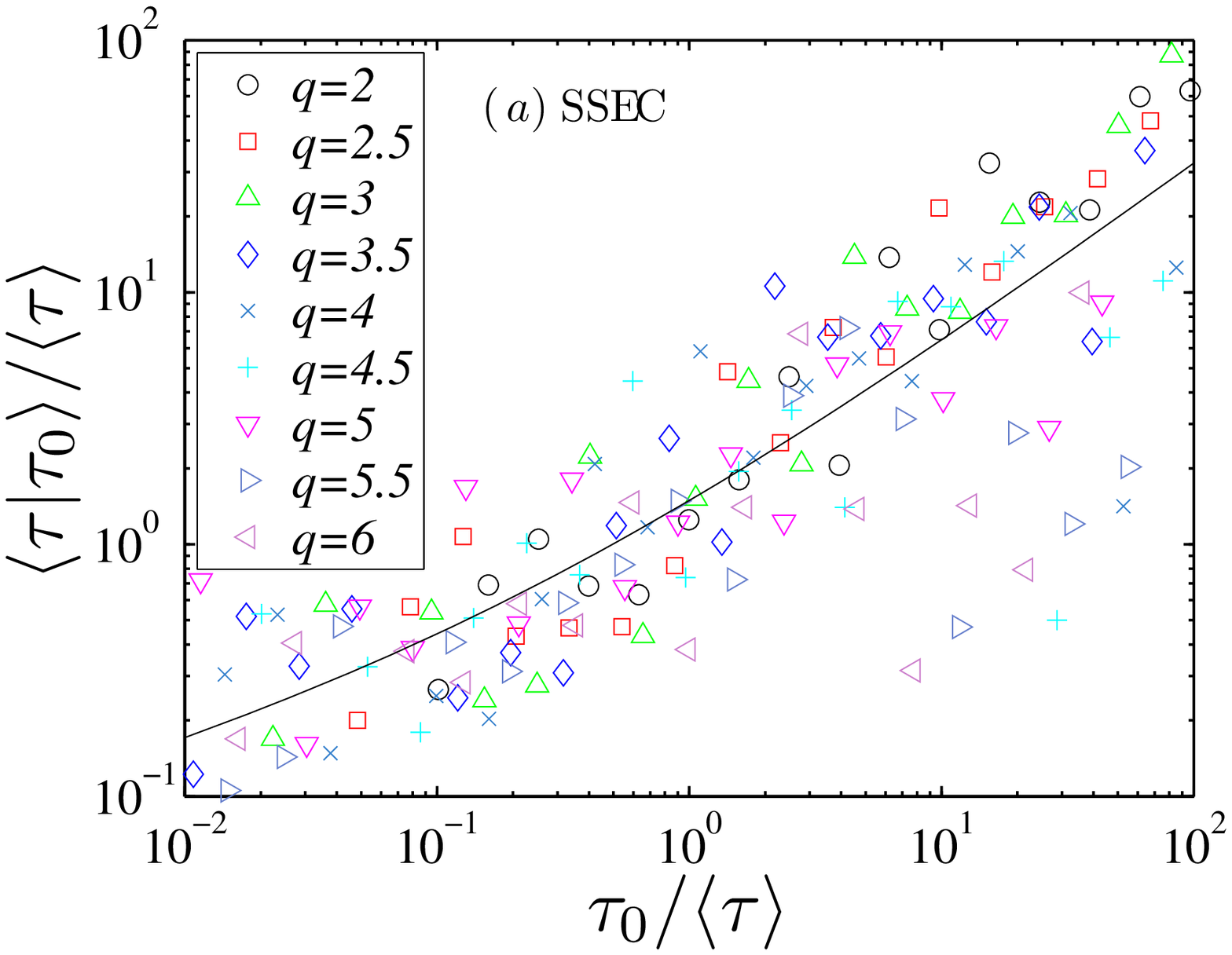}
\includegraphics[width=4cm]{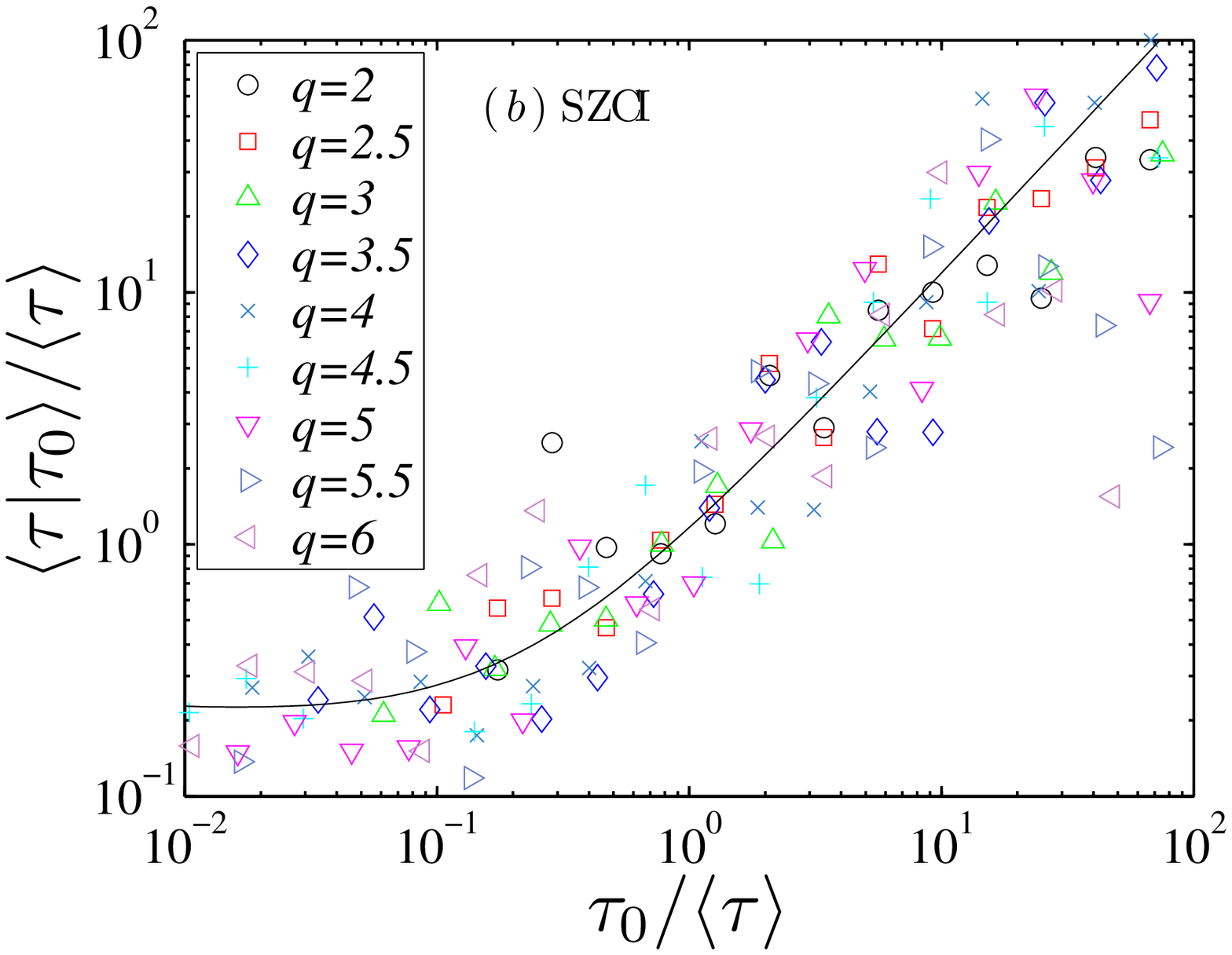}
\includegraphics[width=4cm]{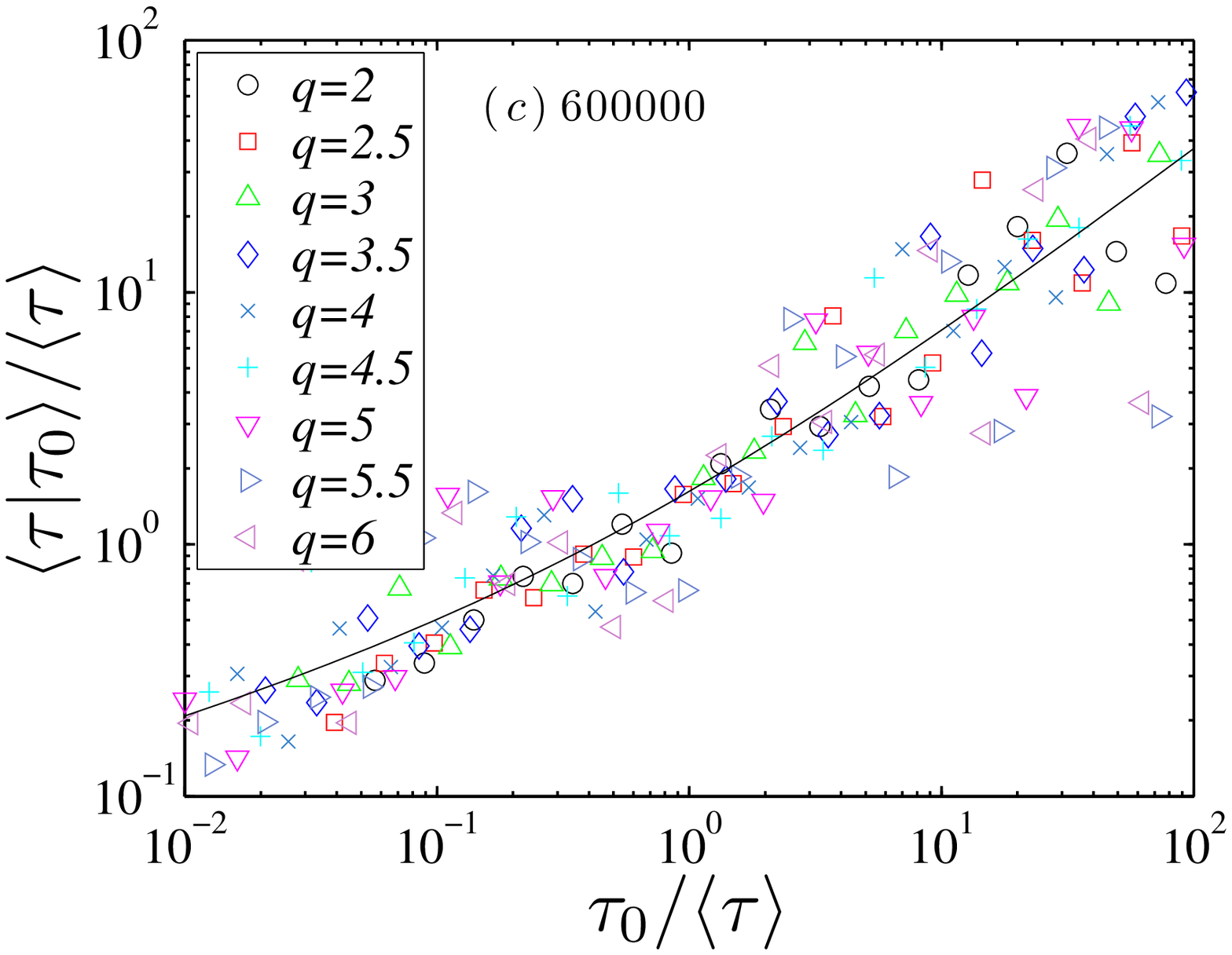}
\includegraphics[width=4cm]{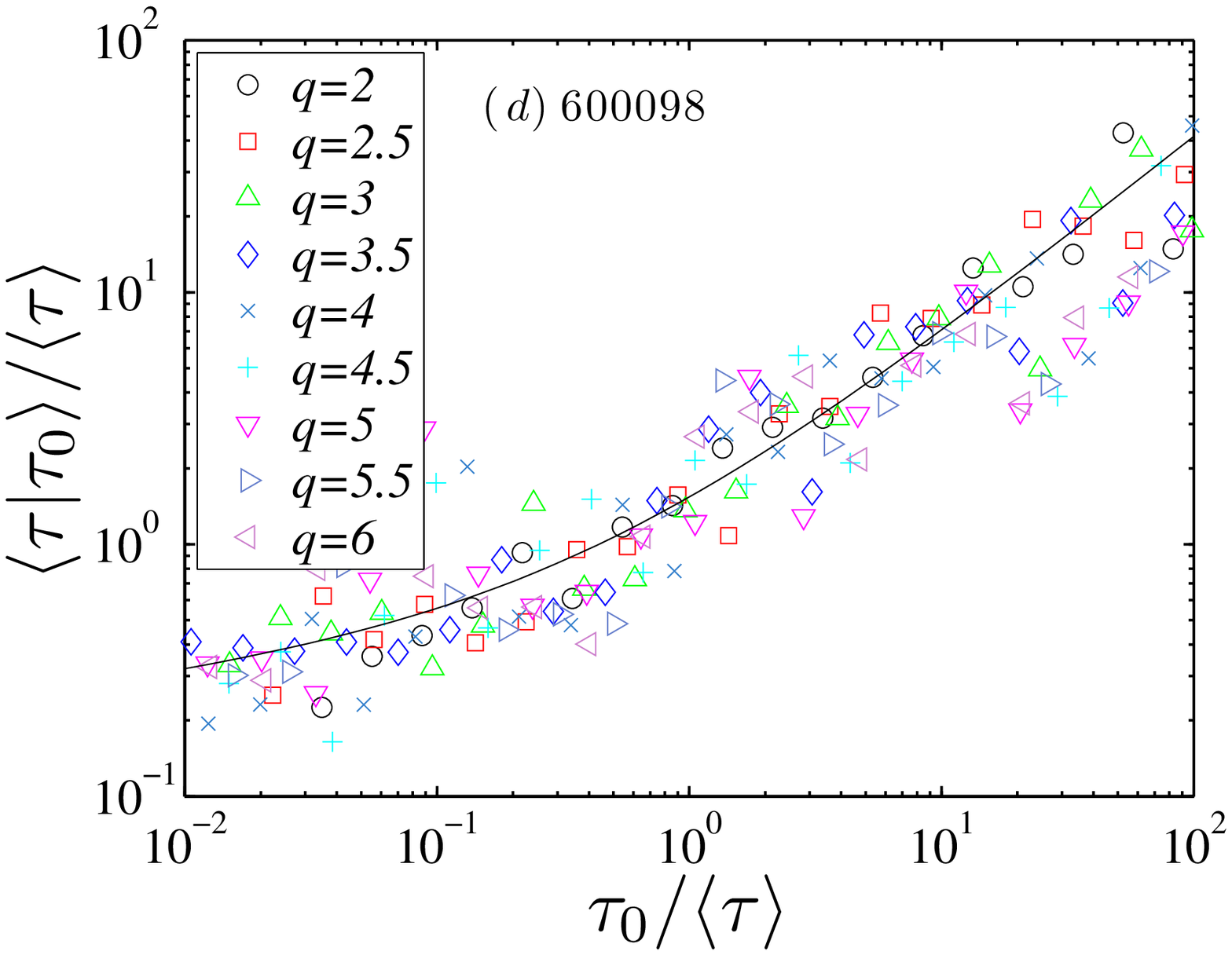}
\includegraphics[width=4cm]{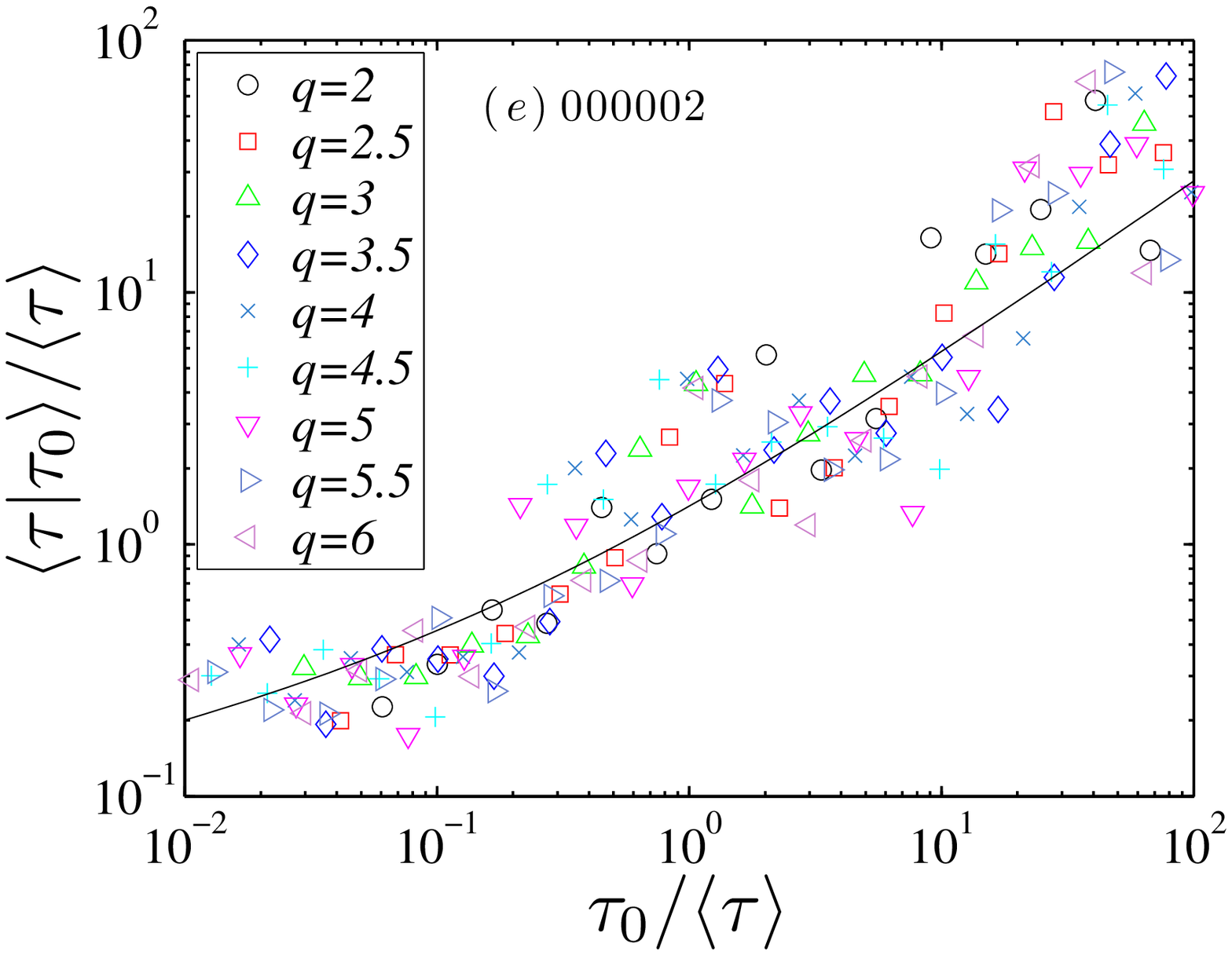}
\includegraphics[width=4cm]{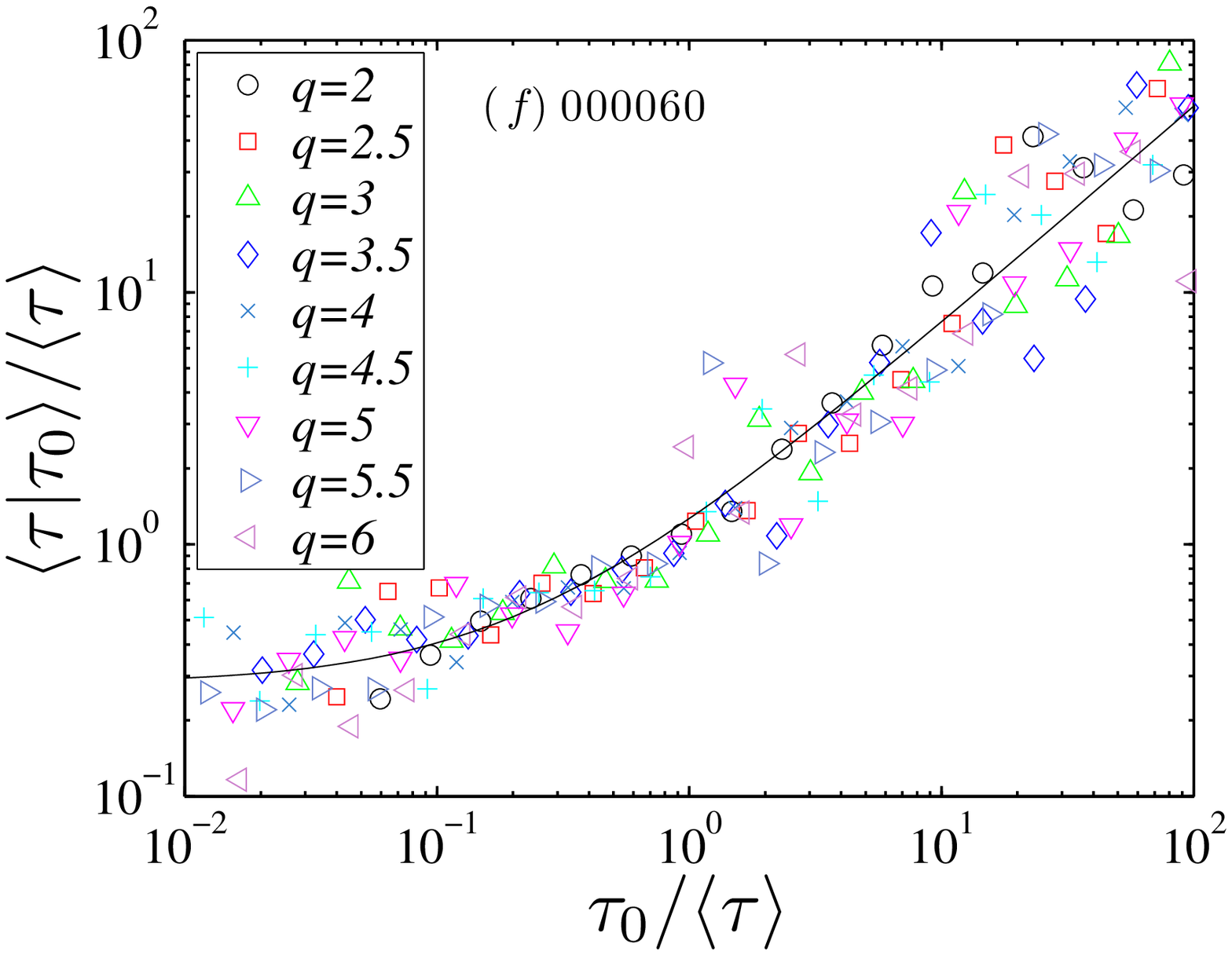}
\caption{\label{Fig:MeanRI1:tau0} (Color online) Scaled mean
conditional return interval $\langle \tau| \tau_0 \rangle/ \langle
\tau \rangle$ as a function of scaled return interval $\tau_0/
\langle \tau \rangle$ for SSEC, SZCI and four representative stocks.}
\end{figure}

To detect the short-term memory of the recurrence intervals, we can
also calculate the mean conditional recurrence interval $\langle \tau | \tau_0 \rangle$
conditioned on the preceding interval $\tau_0$. In Fig.~\ref{Fig:MeanRI1:tau0},
the scaled mean conditional recurrence interval $\langle \tau | \tau_0 \rangle/\langle \tau \rangle$
is plotted with respect to the scaled preceding interval $\tau_0 /\langle \tau \rangle$ for the
two Chinese indices and four representative stocks. Though
$\langle \tau | \tau_0 \rangle/\langle \tau \rangle$ strongly fluctuates
in the whole region of $\tau_0 /\langle \tau \rangle$, it approximately shows a monotonic
increasing tendency as the increase of $\tau_0 /\langle \tau \rangle$. This indicates that
for small (large) preceding interval $\tau_0$ the mean value of the following interval is also small (large),
and this further confirms the short-term memory revealed in the conditional PDF.

\subsection{Detrended fluctuation analysis}
\label{S2:DFA}

To further investigate the long-term memory of the recurrence intervals between trading volumes, we
use the detrended fluctuation analysis (DFA) method, known as a common method of measuring the temporal correlation of a time series
\cite{Peng-Buldyrev-Havlin-Simons-Stanley-Goldberger-1994-PRE,Kantelhardt-Bunde-Rego-Havlin-Bunde-2001-PA,Hu-Ivanov-Chen-Carpena-Stanley-2001-PRE,Chen-Ivanov-Hu-Stanley-2002-PRE,Chen-Hu-Carpena-Bernaola-Galvan-Stanley-Ivanov-2005-PRE}.
The DFA method computes the detrended fluctuation function $F(l)$ of the time series within a window of $l$ data points
after removing a linear trend, and $F(l)$ is expected to follow a scaling form
\begin{equation}
   F(l)\sim l ^ \alpha.
   \label{Eq:DFA}
\end{equation}
The scaling exponent $\alpha$ contains the information of temporal correlation between
the time series: if $\alpha=0.5$ the time series are uncorrelated, and if $0.5<\alpha<1.0$
the time series are long-term correlated. By calculating the exponent $\alpha$ of
the recurrence intervals, we can investigate the temporal
correlation between them.

\begin{figure}[htb]
\centering
\includegraphics[width=4cm]{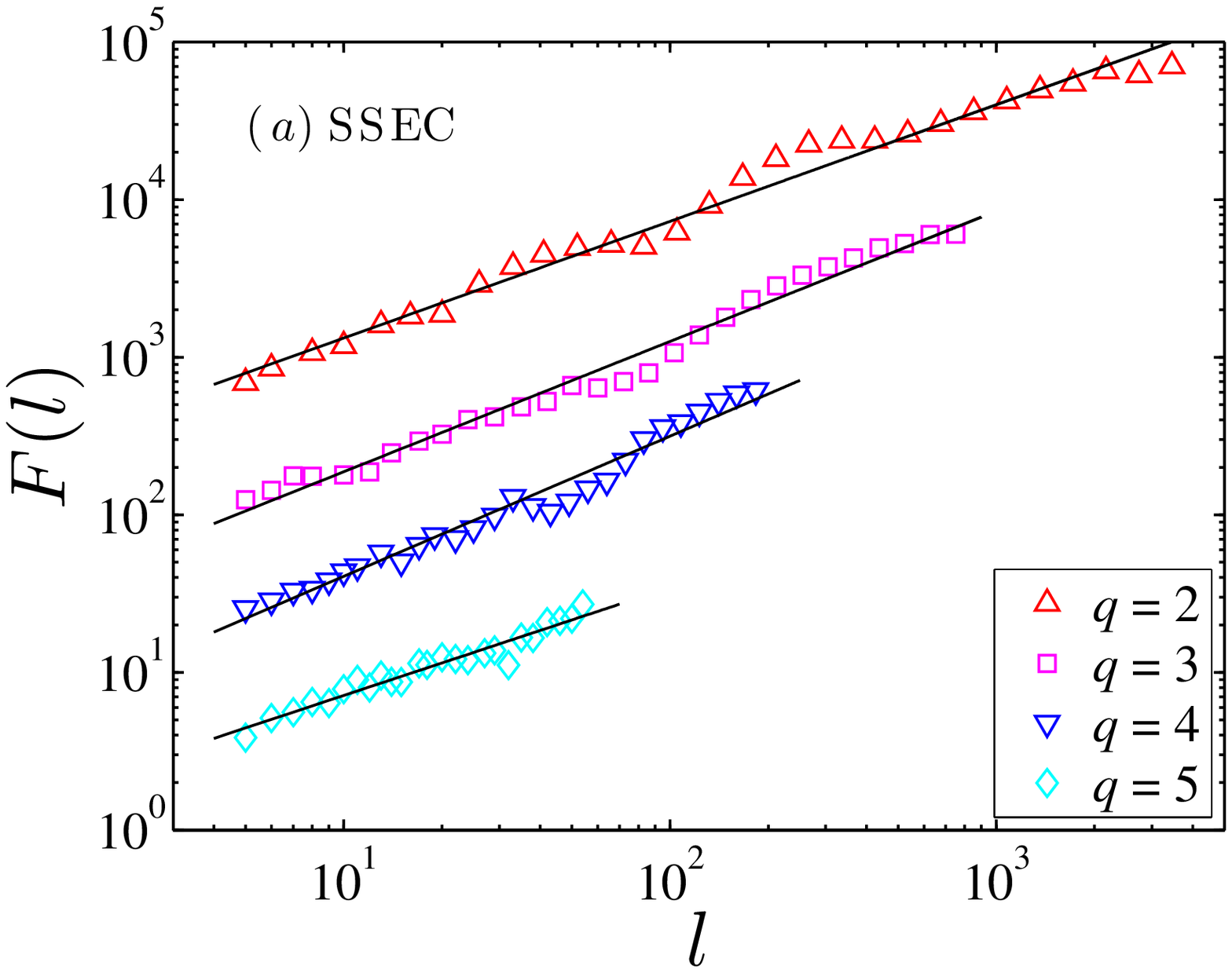}
\includegraphics[width=4cm]{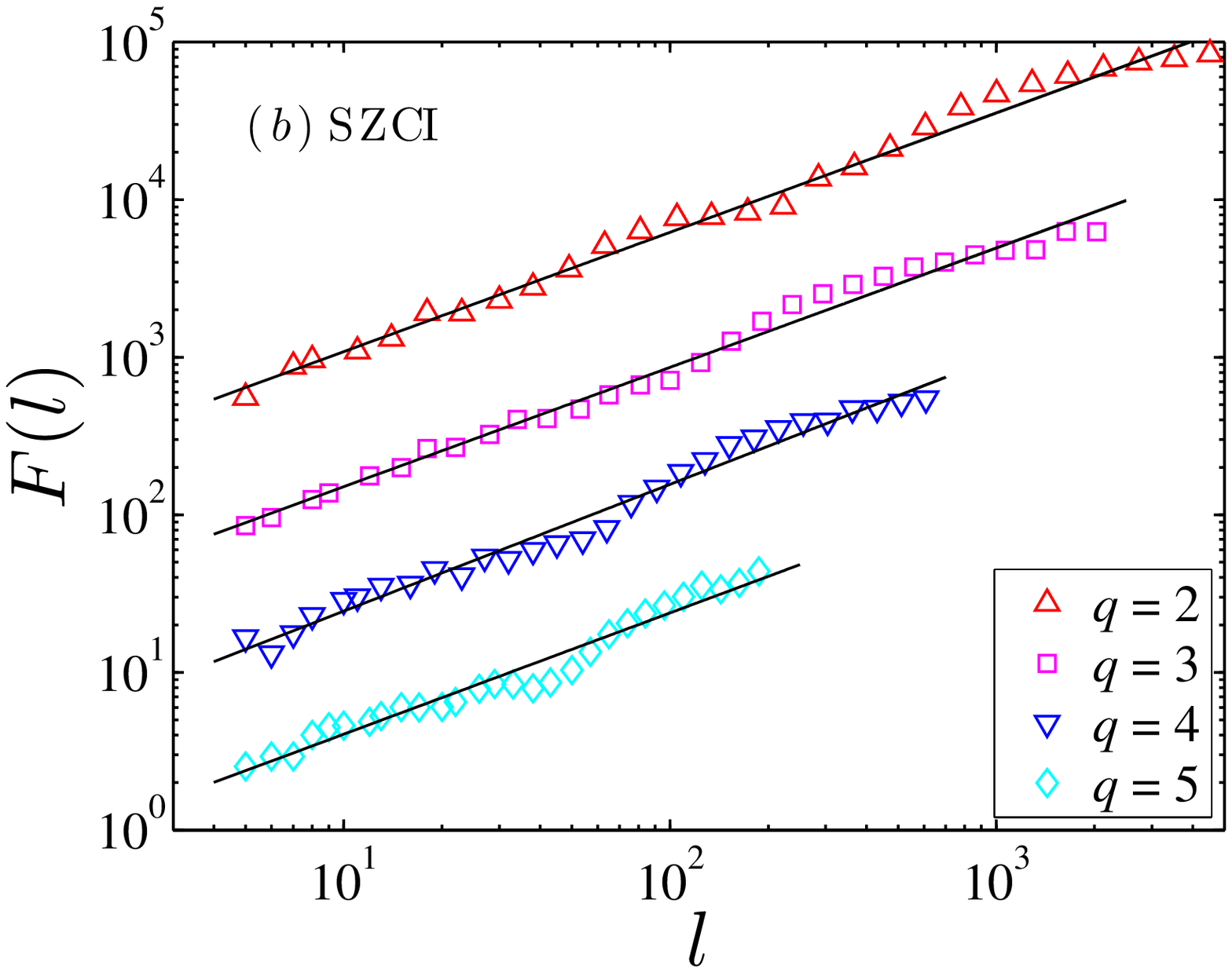}
\includegraphics[width=8cm]{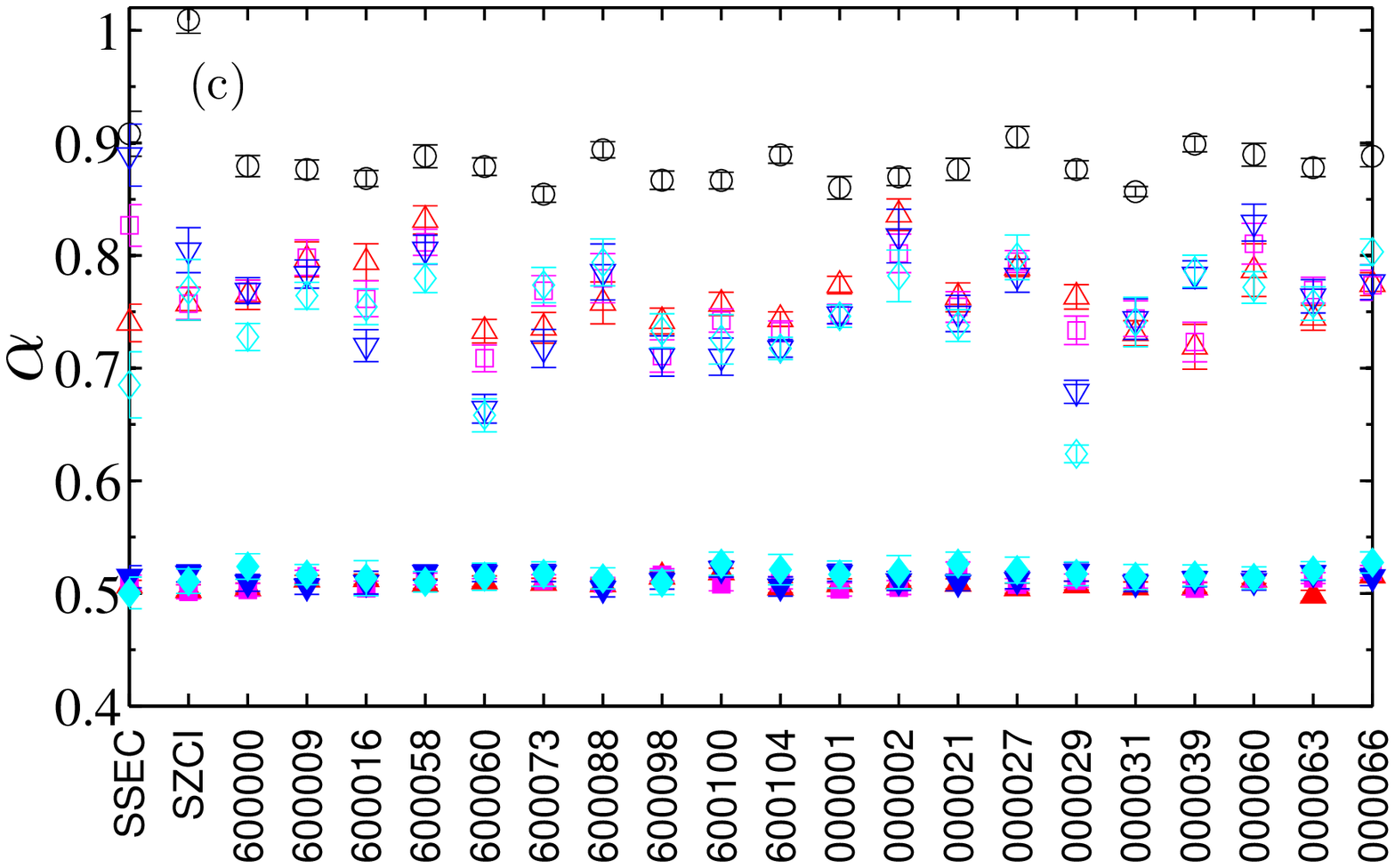}
\caption{\label{Fig:RI:DFA} (Color online) Detrended fluctuation function
$F(l)$ of return intervals for SSEC and SZCI. The curves are
vertically shifted for clarity. Exponents $\alpha$ of trading
volumes (black circles), recurrence intervals (open symbols) and
shuffled recurrence intervals (filled symbols) for two Chinese indices
and $20$ stocks.}
\end{figure}

Fig.~\ref{Fig:RI:DFA} (a) and (b) plot $F(l)$ of the recurrence intervals for the
two Chinese indices. Different symbols represent $F(l)$ for different values of threshold $q$.
One observes that for both two indices $F(l)$ for different $q$ values approximately obey
a scaling form, and similar scaling behaviors are observed for the $20$ individual stocks.
By calculating the slope of its best linear fit in the double logarithmic plot, we can
obtain the estimate of $\alpha$. In Fig.~\ref{Fig:RI:DFA} (c), the estimate
values of $\alpha$ for the two Chinese indices and $20$ stocks are depicted. It is clear
that the values of $\alpha$ for all the indices and stocks are significantly larger than 0.5,
which indicates there exists a long-term memory in the recurrence intervals. We further
assume that the long-term memory of the recurrence intervals may arise from the the long-term
memory of the trading volumes. To verify this we also calculate $F(l)$ of the recurrence intervals
between shuffled trading volumes, in which the long-term memory is artificially eliminated
by shuffling. The exponent $\alpha$ of the recurrence intervals between shuffled trading volumes for the
two Chinese indices and $20$ stocks shows values very close to $0.5$, as illustrated in
Fig.~\ref{Fig:RI:DFA} (c). This may provide an evidence to substantiate the assumption that the long-term
memory of the trading volumes leads to the long-term memory of the recurrence between them.

\section{Dependence of trading volumes on price returns}

\subsection{Correlation between recurrence intervals of trading volumes and their preceding price returns}
\label{S2:Corr:RI:Return}

Empirical studies have shown that the trading volumes are highly
correlated with the price returns, that large price returns are
usually accompanied by large trading volumes. We attempt to
address the question how the price returns affect the trading
volumes based on the return interval analysis. Suppose
$v(t)$ and $v(t+\tau)$ are two consecutive trading volumes exceeding
threshold $q$ and $r(t)$ is the preceding return of
recurrence interval $\tau$ as illustrated in Fig.~\ref{Fig:Cov:RI}, the correlation function is defined as
\begin{equation}
 C(|r|,\tau) = \frac{\langle |r(t)| \tau \rangle - \langle |r(t)| \rangle \langle \tau \rangle}{\sqrt{(\langle |r(t)|^2 \rangle-\langle |r(t)|
\rangle^2)(\langle \tau^2 \rangle-\langle \tau \rangle^2)}}.
 \label{Eq:RetRI:Product}
\end{equation}
It measures the correlation between the magnitude of the price return $|r(t)|$ and the volume
recurrence interval $\tau$ immediately after it. Here $\langle \cdots\rangle$ is the average over time $t$.

\begin{figure}[htb]
\centering
\includegraphics[width=8cm]{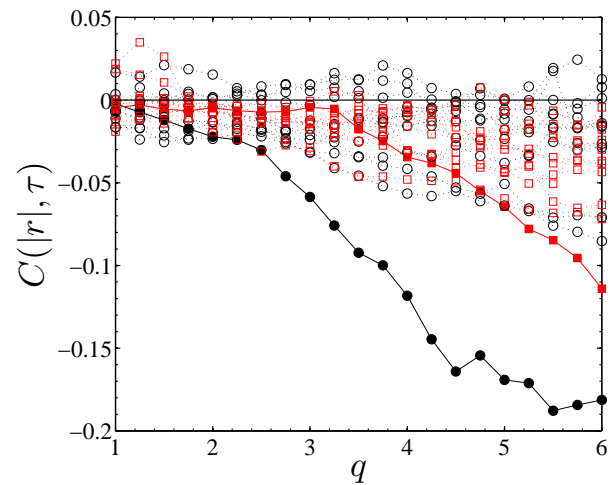}
\caption{\label{Fig:RetRI:Corr} (Color online) Correlation function
between recurrence intervals of trading volumes and their preceding returns for
SSEC (filled black circles), SZCI (filled red squares), $10$ stocks
traded on Shanghai Stock Exchange (open black circles) and $10$
stocks traded on Shenzhen Stock Exchange (open red squares).}
\end{figure}

\begin{figure}[htb]
\centering
\includegraphics[width=8cm]{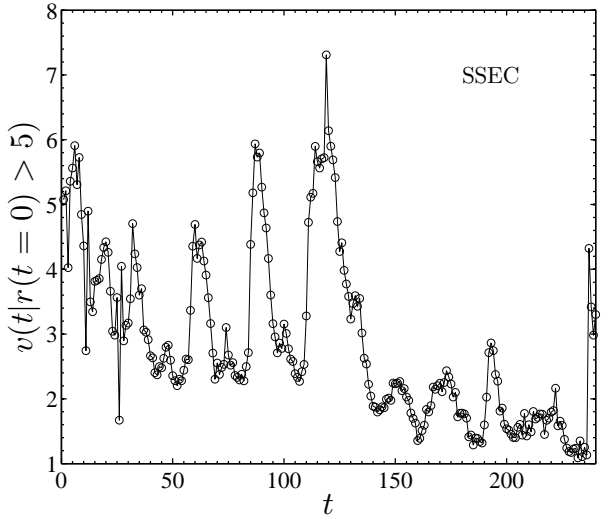}
\caption{\label{Fig:Vol:Evo} (Color online) Evolution of a certain series of normalized trading
volumes conditioned on a large preceding return $r>5$ for SSEC.}
\end{figure}

Fig.~\ref{Fig:RetRI:Corr} plots the correlation function $C(|r|,\tau)$ for the
two Chinese indices and $20$ stocks. For large $q$, $C(|r|,\tau)$ for the two Chinese indices
and most of the individual stocks are significantly smaller than zero. The negative value of $C(|r|,\tau)$
indicates an anti-correlation between the volume recurrence interval and its
preceding return. In other words, once a large positive or negative price change occurs,
the following recurrence interval between large trading volumes tends to show small values,
therefore large trading volumes are more likely to occur after large
price returns. To illustrate this, we randomly select a certain series of trading volumes
initiated with an extremely large return $r(t=0)>5$ for SSEC, and plot it in Fig.~\ref{Fig:Vol:Evo}.
The trading volume strongly fluctuates short after the large return for $t\leq120$, and shows on average larger
values than that after cooling down for a while for $t>120$.

\subsection{Comovement between trading volumes and price returns}
\label{S2:RI:Syn}

\begin{figure}[htb]
\centering
\includegraphics[width=8cm]{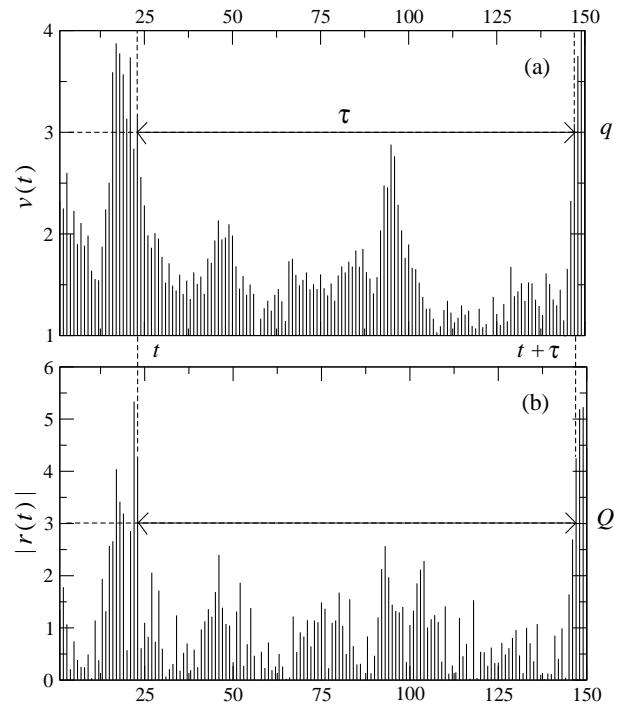}
\caption{\label{Fig:Cov:RI} (Color online) Illustration of normalized trading volumes and absolute price returns which have same recurrence intervals for SSEC.}
\end{figure}

We have shown that there exists an anti-correlation between
the recurrence intervals of trading volumes and their preceding price returns. The evolution curves of trading volumes and absolute price returns are plotted in Fig.~\ref{Fig:Cov:RI}, the observed comovement between them suggests that there might be some relationship between their recurrence intervals. Therefore, we further study the relationship between the recurrences intervals of both trading volumes and price returns. In doing so we calculate the probability that the trading volumes have the same
recurrence intervals as the price returns as illustrated in Fig.~\ref{Fig:Cov:RI}. Suppose two consecutive trading volumes exceeding threshold $q$ occur at time $t$ and $t+\tau$, the probability is calculated as
\begin{equation}
 P(\tau\ |\ |r|>Q) = \frac{N_{\tau,\ |r|>Q}}{N_{\tau}},
 \label{Eq:RI:Syn}
\end{equation}
where $N_{\tau,\ |r|>Q}$ is the number of $\tau$
conditioned on two consecutive absolute returns exceeding the threshold $Q$, i.e., $|r(t)|>Q$ and $|r(t+\tau)|>Q$, and
$N_{\tau}$ is the total number of $\tau$.

\begin{figure}[htb]
\centering
\includegraphics[width=4cm]{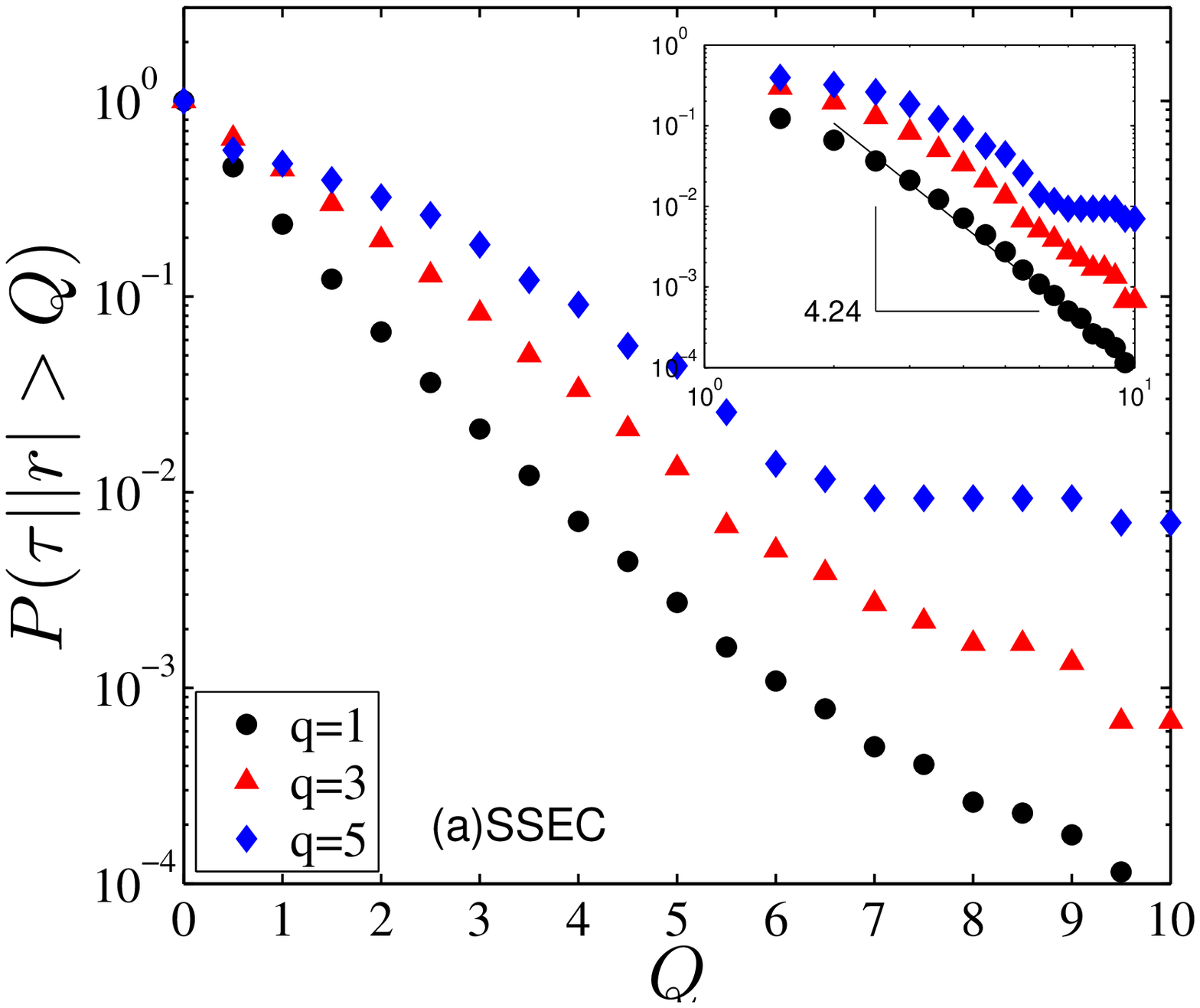}
\includegraphics[width=4cm]{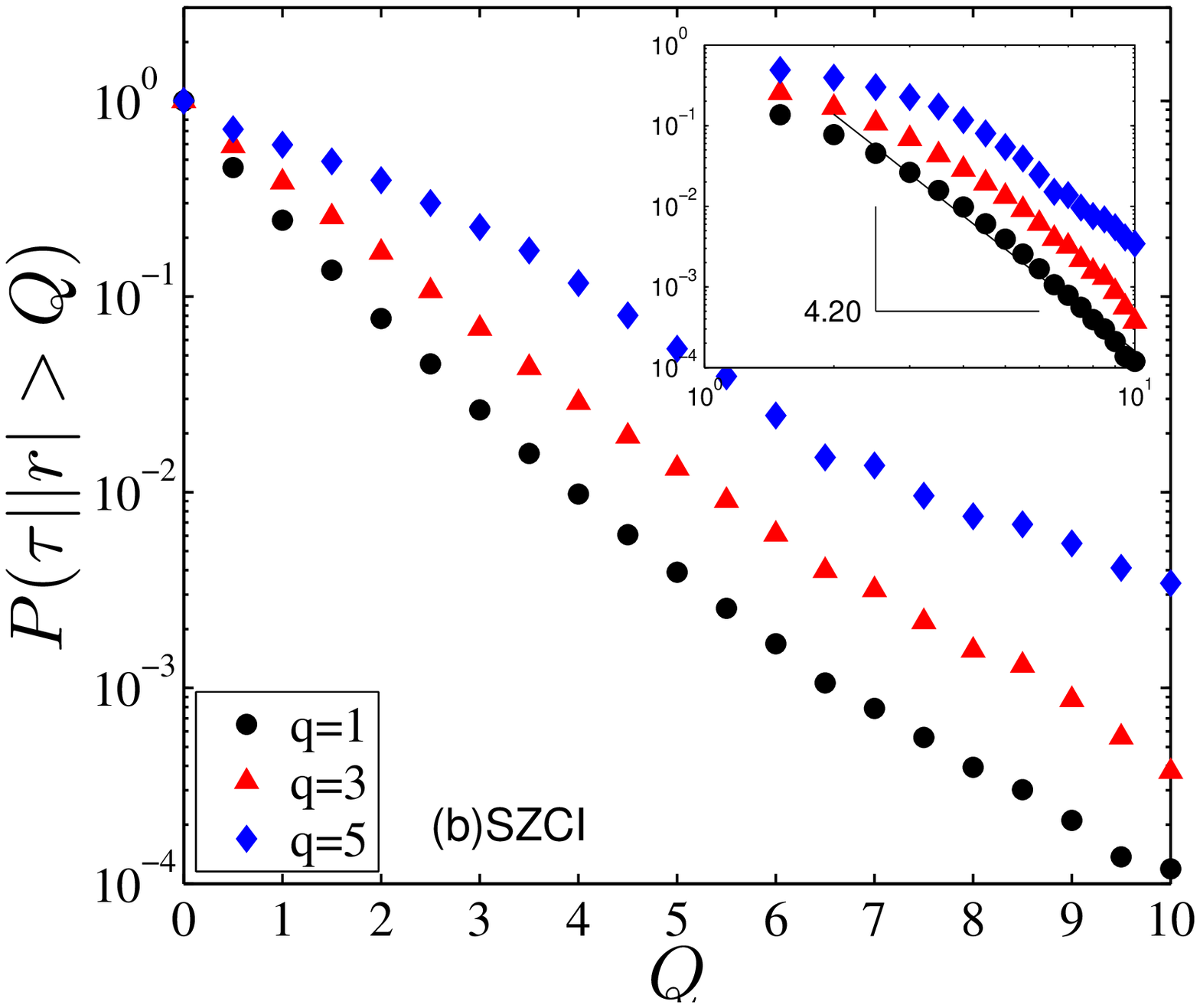}
\includegraphics[width=4cm]{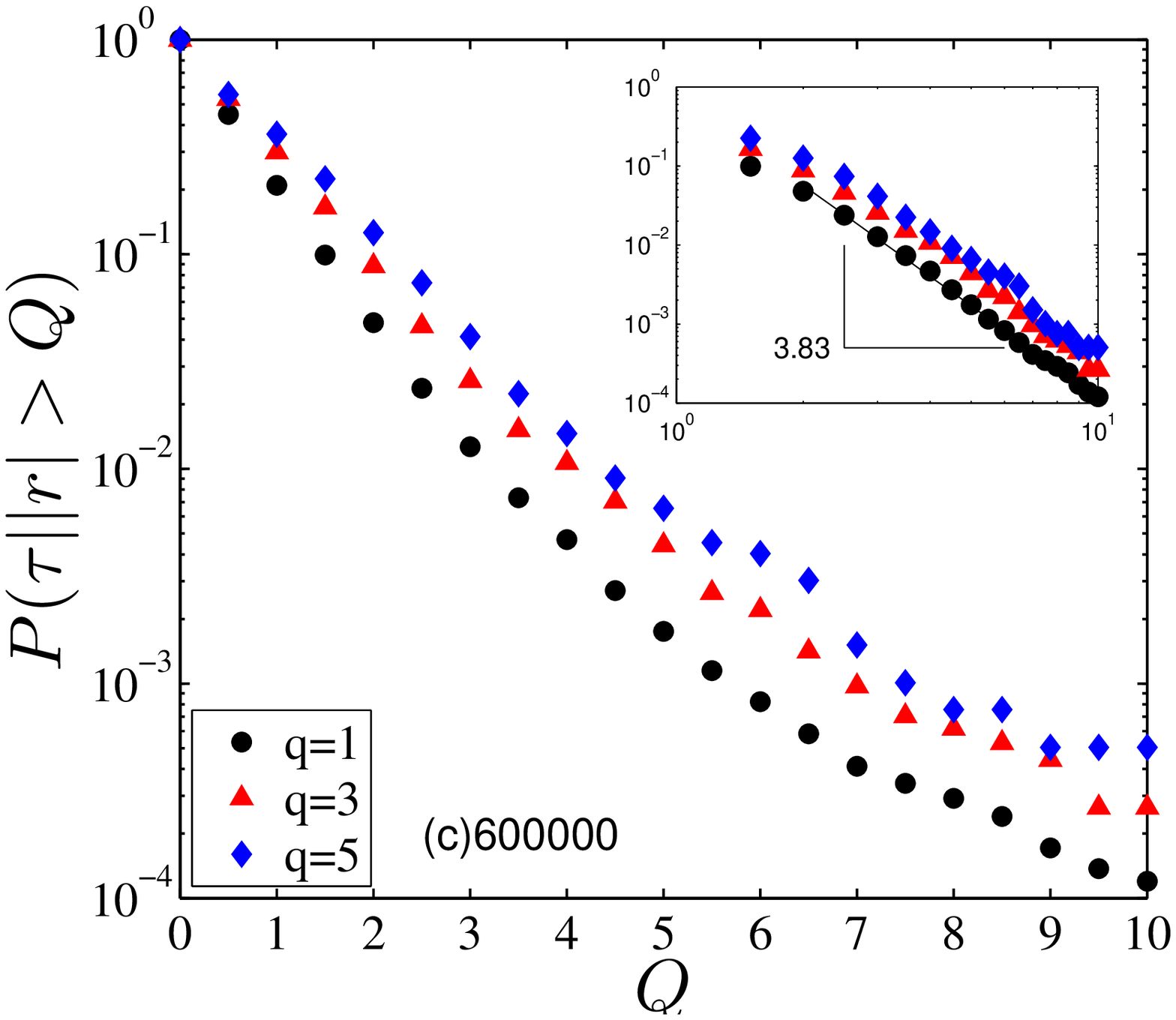}
\includegraphics[width=4cm]{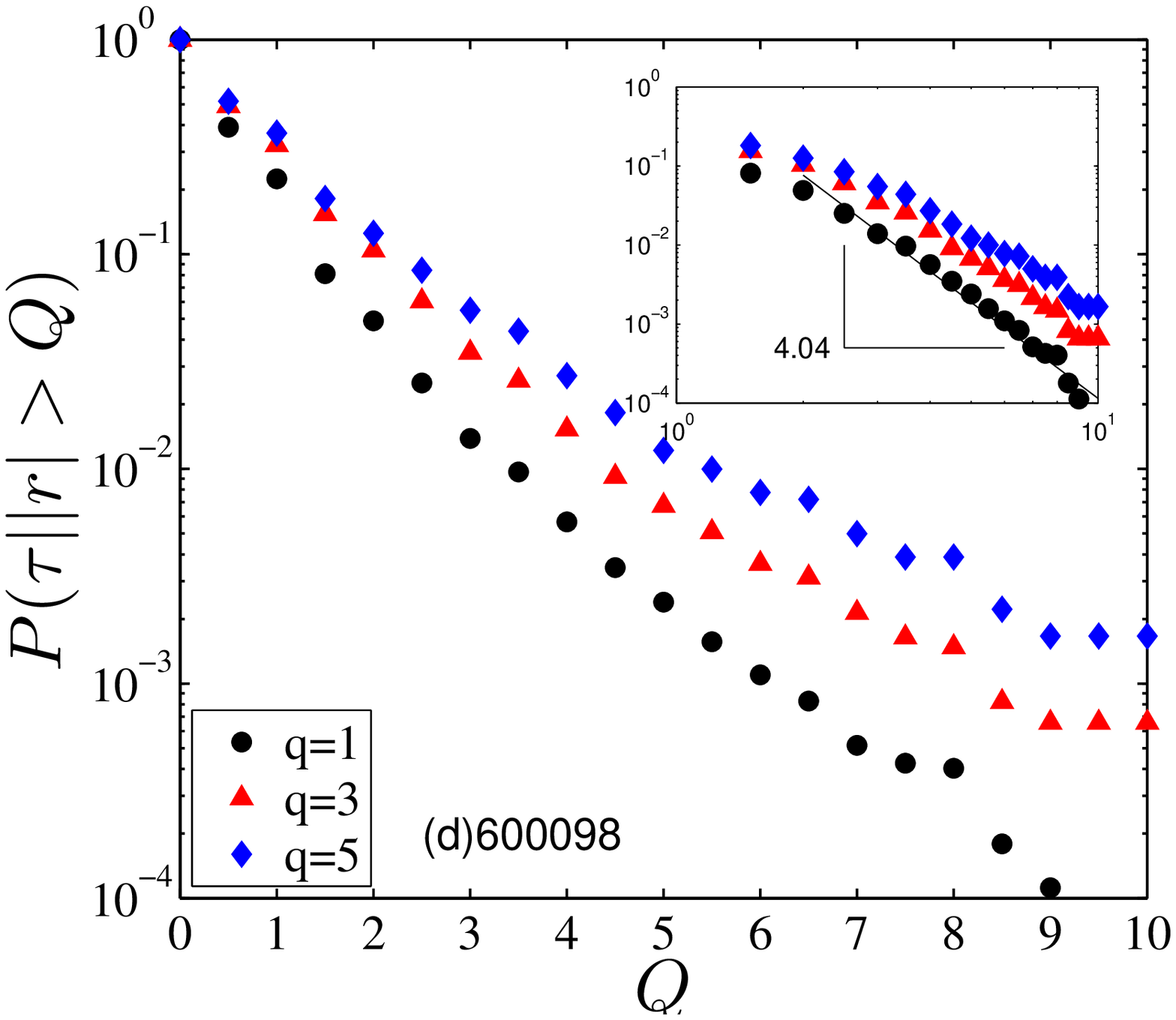}
\includegraphics[width=4cm]{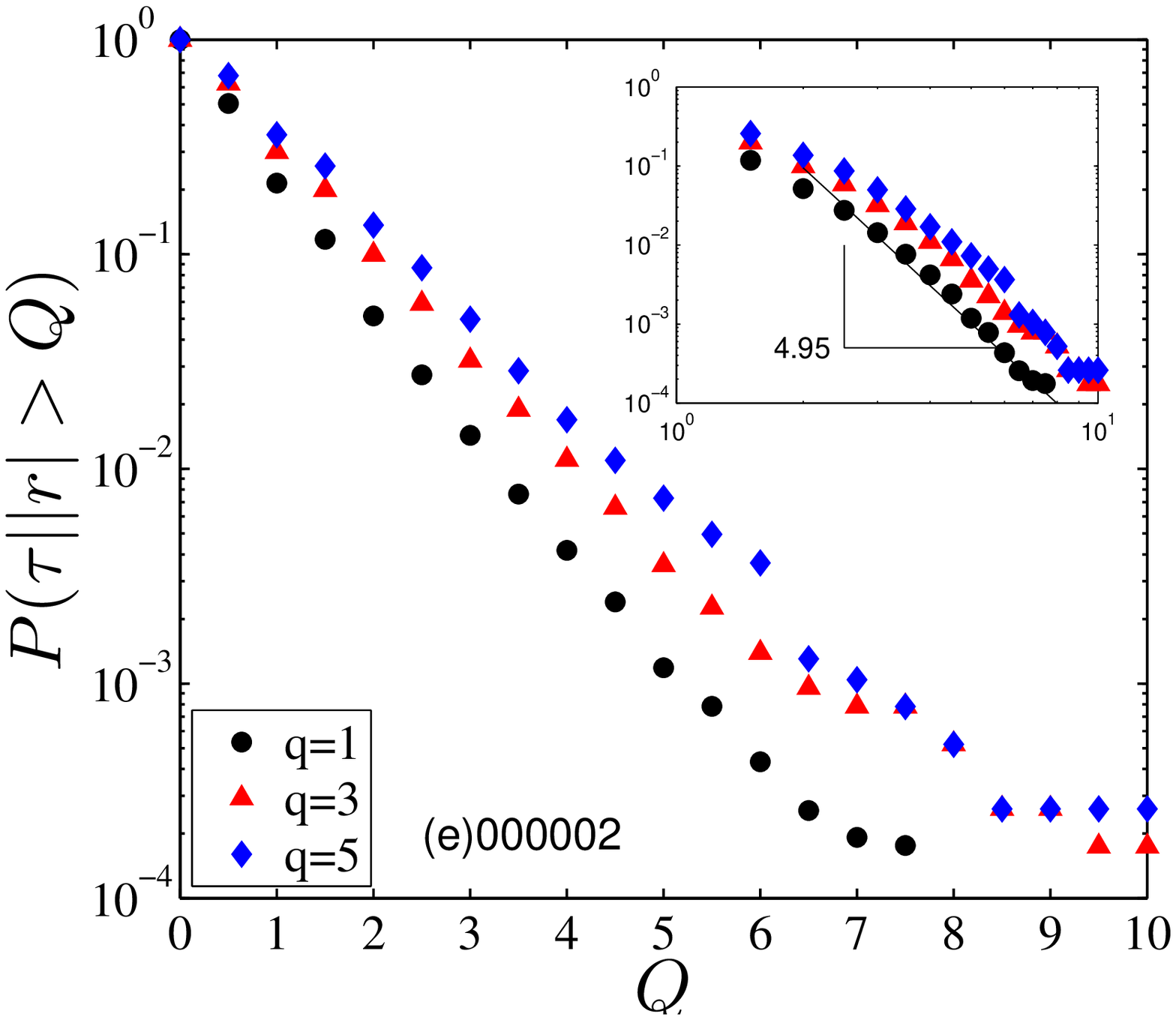}
\includegraphics[width=4cm]{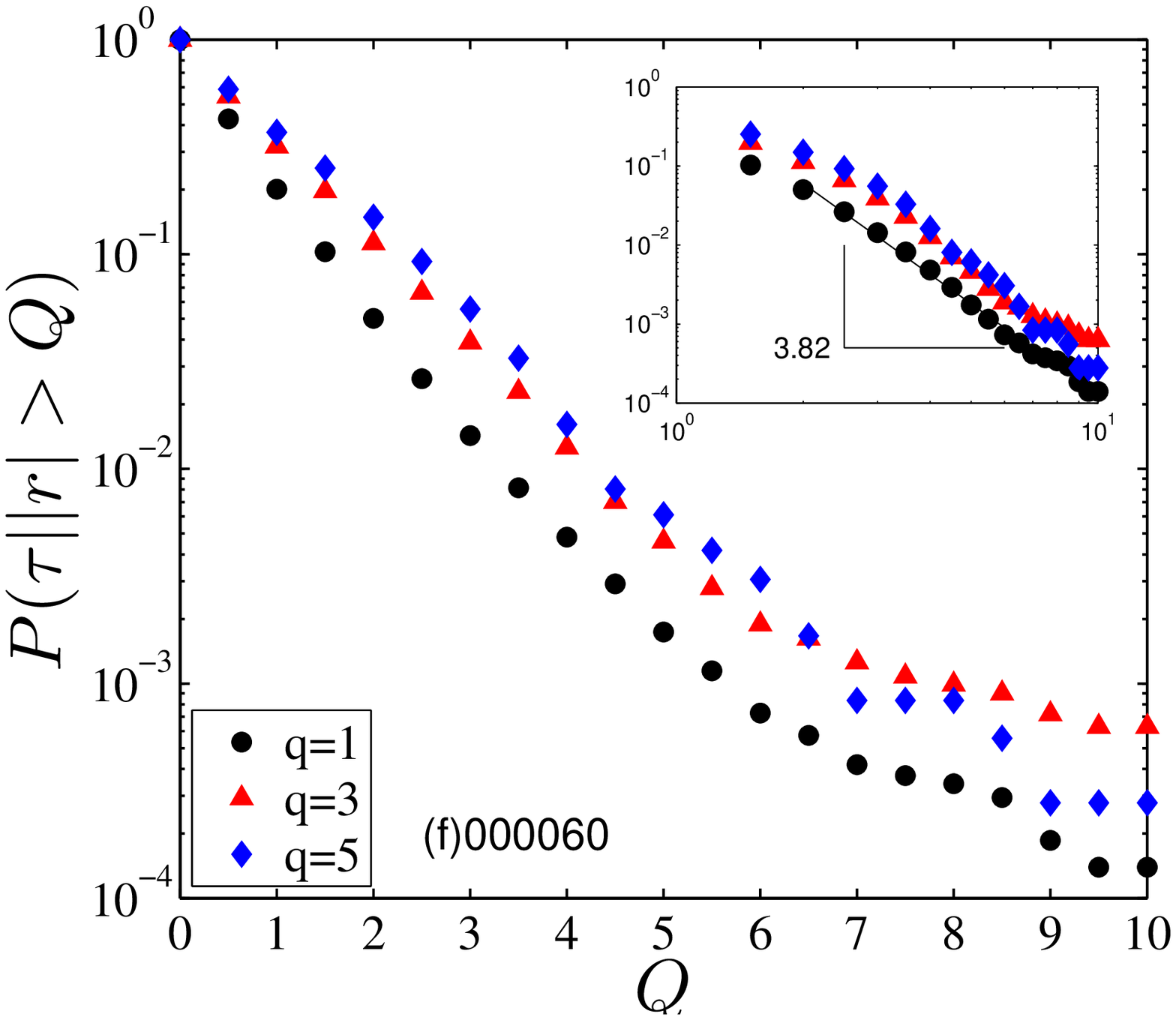}
\caption{\label{Fig:RI:Syn} (Color online) Probability of volume
intervals conditioned on large volatilities $|r|>Q$ for SSEC, SZCI
and four representative stocks. }
\end{figure}

In Fig.~\ref{Fig:RI:Syn}, $P(\tau\ |\ |r|>Q)$ is plotted as a function of
the threshold $Q$ for the two Chinese indices and four representative
stocks. In principle, $P(\tau\ |\ |r|>Q)$ equals one
when $Q=0$, and approximately follows a power-law decay for large scales of $Q$ as
shown in the insets in Fig.~\ref{Fig:RI:Syn}. One observes that
$P(\tau\ |\ |r|>Q)$ for large $q$ is obviously
larger than that for small $q$, displaying a power-law
tail with an exponent similar to that for small $q$. This
indicates that the comovement between trading volumes and price
returns is more pronounced for large thresholds $q$ and $Q$,
which further confirms the close relationship between the trading
volumes and the price returns.

\section{Conclusion}
\label{S1:Conclusion}

In this paper, we have introduced the recurrence interval analysis to the study of stock market trading volumes. The recurrence intervals between the trading volumes above a threshold $q$ for the two Chinese indices and 20 liquid Chinese stocks are carefully studied. Though the scaled PDFs of the recurrence intervals for different $q$ values diverge slightly for small intervals, they tend to show scaling behavior for large intervals. The goodness of fit tests are further performed to find out the specific form of the scaling function, and the results demonstrate that it could be approximated by a power law. The measurements of the conditional PDF, the mean conditional recurrence interval and the DFA method are carried out to detect the short-term and long-term memory of the recurrence intervals, and both memory effects are clearly observed.

The study of market impact of trading volumes is supposed to be of great importance for both theoretical and practical reasons. Here, we tried to study the correlation between trading volumes and price returns using the recurrence interval analysis method. We calculated the correlation between the recurrence intervals of trading volumes and their proceeding price returns, and found that the trading volumes following large price returns are more likely to show large values, which may corresponds to the unstable state when after a market burst. The pronounced comovement between large trading volumes and large price returns further confirms our previous findings that large price returns are usually accompanied by large trading volumes. In fact, our study mainly focuses on the relationship between trading volumes and the magnitude of price returns. Further study concerning the dependence of trading volumes on price returns taking into account the effect of the signs of price returns needs to be proceeded, and this will certainly help to study the risk estimation in financial markets.

\begin{acknowledgments}
We are grateful to Kun Guo (Research Center on Fictitious Economics
\& Data Science, Chinese Academy of Sciences) for retrieving the
data analyzed in this work and Gao-Feng Gu (School of Business, East
China University of Science and Technology) for preprocessing the
data. This work was partially supported by ``Chen Guang'' project and ``Shu
Guang'' project sponsored by Shanghai Municipal Education Commission
and Shanghai Education Development Foundation (Nos. 2008CG37 and
2008SG29), the Program for New Century Excellent Talents in
University (No. NCET-07-0288), and the National Natural Science
Foundation (No. 10905023).
\end{acknowledgments}

\bibliography{E:/Papers/Auxiliary/Bibliography} 

\end{document}